\documentclass[12pt,preprint]{aastex}
\usepackage{natbib}
\usepackage{color}
\usepackage{amsmath}
\usepackage[colorlinks,linkcolor=blue,anchorcolor=blue,citecolor=blue]{hyperref}

\begin{document}

\title{QUANTIFYING THE TOPOLOGY AND EVOLUTION OF A MAGNETIC FLUX ROPE ASSOCIATED WITH MULTI-FLARE ACTIVITIES}
\author{\sc{Kai Yang$^{1,2}$, Yang Guo$^{1,2}$, M. D. Ding$^{1,2}$}}
\affil{$^1$School of Astronomy and Space Science, Nanjing University, Nanjing 210023, China} \email{dmd@nju.edu.cn}
\affil{$^2$Key Laboratory for Modern Astronomy and Astrophysics (Nanjing University), Ministry of Education, Nanjing 210023, China}

\begin{abstract}
Magnetic flux rope (MFR) plays an important role in solar activities.
A quantitative assessment of the topology of an MFR and its evolution is crucial for a better understanding of the relationship between the MFR and the associated activities.
In this paper, we investigate the magnetic field of active region 12017 from 2014 March 28 to 29, where 12 flares were triggered by the intermittent eruptions of a filament (either successful or confined).
Using the vector magnetic field data from the Helioseismic and Magnetic Imager on board the \textit{Solar Dynamics Observatory}, we calculate the magnetic energy and helicity injection in the active region, and extrapolate the 3D magnetic field with a nonlinear force-free field model.
From the extrapolations, we find an MFR that is cospatial with the filament.
We further determine the configuration of this MFR by a closed quasi-separatrix layer (QSL) around it.
Then, we calculate the twist number and the magnetic helicity for the field lines composing the MFR.
The results show that the closed QSL structure surrounding the MFR gets smaller as a consequence of the flare occurrence.
We also find that the flares in our sample are mainly triggered by kink instability.
Moreover, the twist number varies more sensitively than other parameters to the occurrence of flares.
\end{abstract}
\keywords{Sun: flares --- Sun: magnetic topology --- Sun: flares}


\section{INTRODUCTION}
It is widely accepted that magnetic flux rope (MFR) represents the core field of an active region and plays an important role in triggering eruptive events, like solar flares and coronal mass ejections (CMEs) \citep{2000Amari,2014Amari,2005Fan,2010Aulanier,2012Zhang,2012Cheng,2013Cheng,a2014Cheng,b2014Cheng,2010Guo,2012Guo}.
In observations, the MFR usually appears as a hot channel structure in extreme-ultraviolet \citep{2012Cheng,2013Cheng,a2014Cheng,b2014Cheng} and a prominence could correspond to the lower part of it \citep{1999Aulanier}.
Basically, its structure is a group of highly twisted magnetic field lines in the solar corona.
Quantifying the topological quantities of an MFR is thus important for diagnosing its activity level.
The most relevant quantities include the relative magnetic helicity, twist number, and quasi-separatrix layer (QSL).

The magnetic helicity describes the magnetic field complexity including the twist, writhe, knot, and linkages of magnetic field lines.
Globally, magnetic helicity is a conserved quantity in ideal magnetohydrodynamics (MHD).
In resistive MHD, this quantity is almost conserved as what was predicted by \cite{1974Taylor}.
Since magnetic helicity has the property of inverse cascade from small scales to large scales \citep{1975Frisch,1976Pouquet}, it can avoid the dissipation in small scales, where the magnetic Reynolds number is large.
Therefore, variation of the magnetic helicity in a 3D volume is just a result of the helicity flux flowing into and out of the boundaries.
From the definition of the magnetic helicity, $H = \int_{V}\mathbf{A\cdot B}\rm d \mathcal{V}$, it is gauge invariant in a closed space where the magnetic field lines are confined.
However, this conservation property is not valid in an open space, where the field lines can go out through the boundaries, in which the magnetic helicity may change in quantity with different gauges.
This problem has been resolved by invoking the concept of the relative magnetic helicity \citep{1984Berger}, which is defined as $H_R = \int_{V} (\mathbf{A}+\mathbf{A}_R)\cdot(\mathbf{B}-\mathbf{B}_R)\rm d \mathcal{V}$.
Here, the field $\mathbf{B}_R$ is a reference magnetic field and $\mathbf{A}_R$ is the corresponding vector potential.
The reference field should have the same normal field component as that of the real magnetic field on the boundaries.
Thus, for convenience, the potential field $\mathbf{B}_p$ is always chosen as the reference field.

Magnetic helicity injection into solar active regions has been studied in a large number of papers \citep{2008Park,a2010Park,b2010Park,2013Park,2012Tziotziou,2012Jing,2013Tziotziou,2014Liu}.
\cite{2008Park} found a two-step magnetic helicity injection before major flares: a monotonically increasing phase and a following nearly constant injection phase.
More quantitatively, \cite{b2010Park} revealed that the magnetic helicity injection rate in flaring ARs is two orders of magnitude higher than that in nonflaring ARs.
Magnetic helicity injection into an existing active region with opposite helicity sign may trigger flare events \citep{a2010Park,2013Park}.
\cite{2012Jing} found that magnetic helicity appears to increase or decrease prior to flares.
\cite{2012Tziotziou,2013Tziotziou} proposed a monotonic correlation between free energy and relative magnetic helicity.
Based on a theoretical research, \cite{2006Zhang} suggested that there would be an upper bound for the total magnetic helicity in the force-free field.

Along with the magnetic helicity injection into the solar corona, free magnetic energy is always accumulated simultaneously, and the magnetic field gradually departs from the potential field.
Some typical structures characterizing the activity level may ensue, such as an MFR.
In particular, \cite{2013Guo} found a quantitative relationship between  the helicity injection and the twist number of the MFR.
If the twist number increases to some critical value, the kink instability would occur \citep{1979Hood,1981Hood,2004Torok,2005Fan}.
A recent study by \cite{2016LiuR} showed that the twist tends to increase and decrease before and after the flares, respectively.
On the other hand, if the decay index of the background field, in which the MFR is embedded, is larger than some critical value, the torus instability can occur \citep{2006Kliem,2010Aulanier,2010Demoulin}.
Note that, however, there are some differences in the critical value from different models \citep{2010Demoulin,2010Olmedo} and laboratory experiments \citep{2015Myers}.

Furthermore, the structure of the magnetic field is usually characterized by the QSL, which denotes the place where the magnetic field line connectivity changes dramatically \citep{1995Priest,a1996Demoulin,b1996Demoulin,1997Demoulin,b2006Demoulin,1997Mandrini,2002Titov}.
The QSL can be visualized by calculating the squashing factor, $Q$, which measures the gradient of the field line linkage \citep{2002Titov}.
Many studies have shown that magnetic reconnection can favourable take place at the QSL \citep{1999Milano,2005Aulanier,2009Lawrence,2010WilmotSmith,2012Gekelman}, leading to occurrence of solar flares \citep{1997Demoulin,2012Savcheva,2013Janvier,2013Guo,2015Yang}.

Although a qualitative relationship between MFRs and solar eruptions has likely been established, more quantitative assessment of the key parameters including the QSL, twist number, magnetic helicity, and their temporal evolutions is still needed for a better understanding of the occurrence condition of solar eruptions.
For this purpose, we select the well observed NOAA AR 12017, which contains an MFR and produces multi-flare activities, for the current research.
To our knowledge, among the available data of the same kind, this AR is unique in that it produces more than 10 flares within two days (2014 March 28-29) along with the intermittent eruptions (either successful or confined) of an MFR. 
From the \textit{GOES} soft X-ray flare list, there are eleven C-class, one M-class, and one X-class flares that occurred in this active region.
The X-class flare has been studied by other authors in many aspects.
\cite{2014Judge} reported the sunquake associated with this X-class flare.
\cite{2015Kleint} found that the acceleration of the filament eruption that leads to this flare is as fast as $\sim$ 3--5 km s$^{-2}$.
\cite{2015Aschwanden} evaluated the energy dissipation to be about $(29\pm3)\times10^{30}$ erg during the flare.
\cite{2015LiuC} proposed a two phase process to explain the asymmetric filament eruption that trigger this flare.
The data from the recently launched Interface Region Imaging Spectrograph have also been used in the study of this flare.
For example, \cite{2015Young} studied the post-flare loops and \cite{2015Li} made an analysis of the chromospheric evaporation during the flare.
However, the above researches are mostly concentrated to a single event, without paying attention to the whole evolution of the active region. In this paper, we focus on the evolution of the MFR during the whole period of time when the series of flares occurred.

The purpose of this paper is to explore how the evolution of the MFR is related to the flares. 
To obtain the 3D magnetic field of the volume containing the MFR, we apply the extrapolation technique to a time series of magnetograms with a 12 minute cadence.
The MFR can then be defined and its twist number and relative magnetic helicity can be quantitatively calculated.
This paper is organized as follows.
The extreme-ultraviolet observations and the magnetic field evolution are introduced in Section \ref{Sect.2}.
Section \ref{Sect.3} describes the magnetic field analysis and results.
In particular, we discuss the method for energy and helicity injection in Section \ref{Sect.3.1}, the magnetic field extrapolation in Section \ref{Sect.3.2} and the quantitive results of the MFR and the ambient magnetic field in Sections \ref{Sect.3.3}--\ref{Sect.3.5}.
Section \ref{Sect.4} gives a discussion and summary.

\section{OBSERVATIONS} \label{Sect.2}
\subsection{EUV Observations} \label{Sect.2.1}
\begin{figure}
\centering
\includegraphics[width=1.0\textwidth]{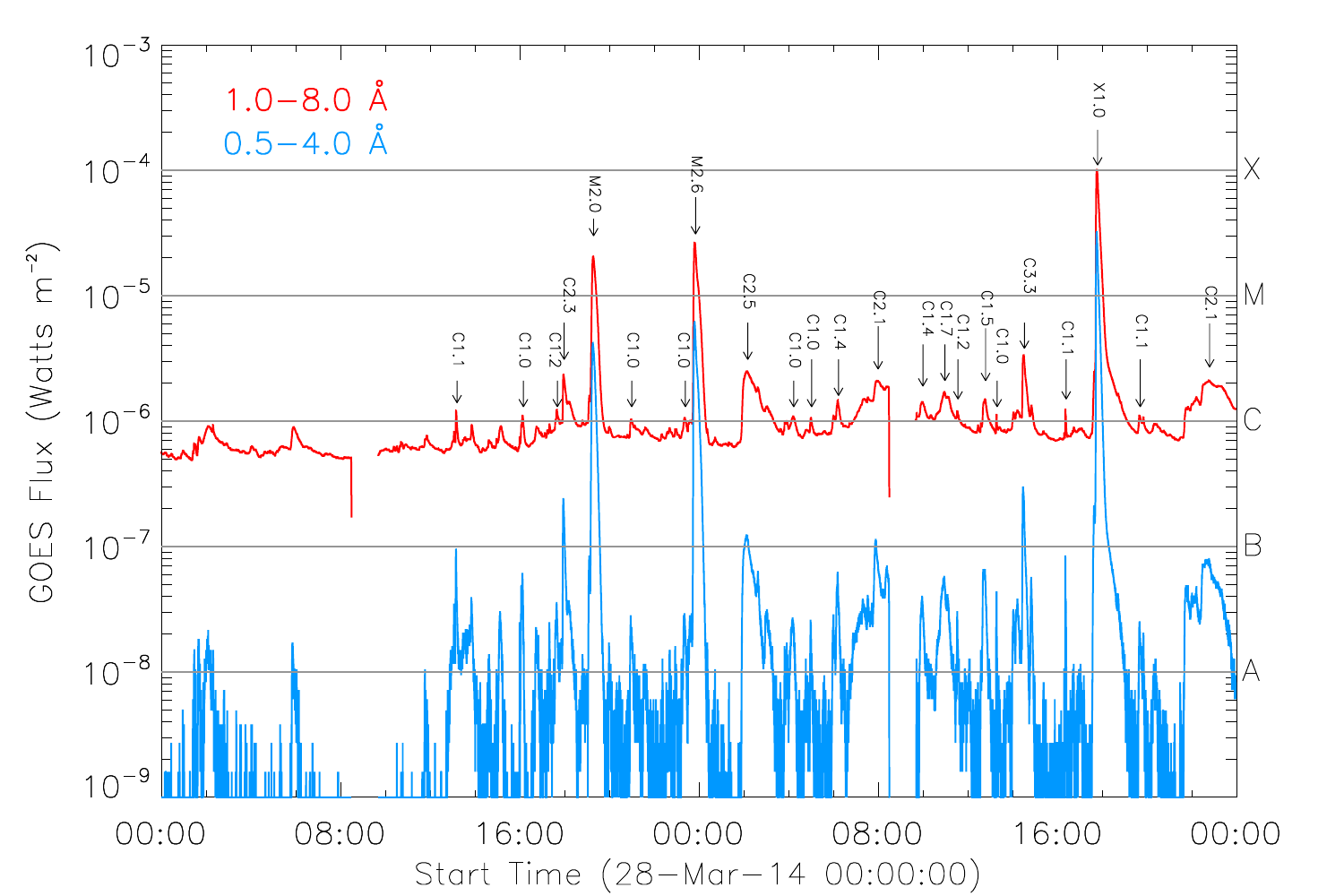} 
\caption{\textit{GOES} soft X-ray flux from 2014 March 28 00:00 UT to March 30 00:00 UT. Red and blue curves represent fluxes in 1.0--8.0 \AA \ and 0.5--4.0 \AA , respectively.}
\label{fig:1}
\end{figure}

\begin{figure}
\centering
\includegraphics[width=1.0\textwidth]{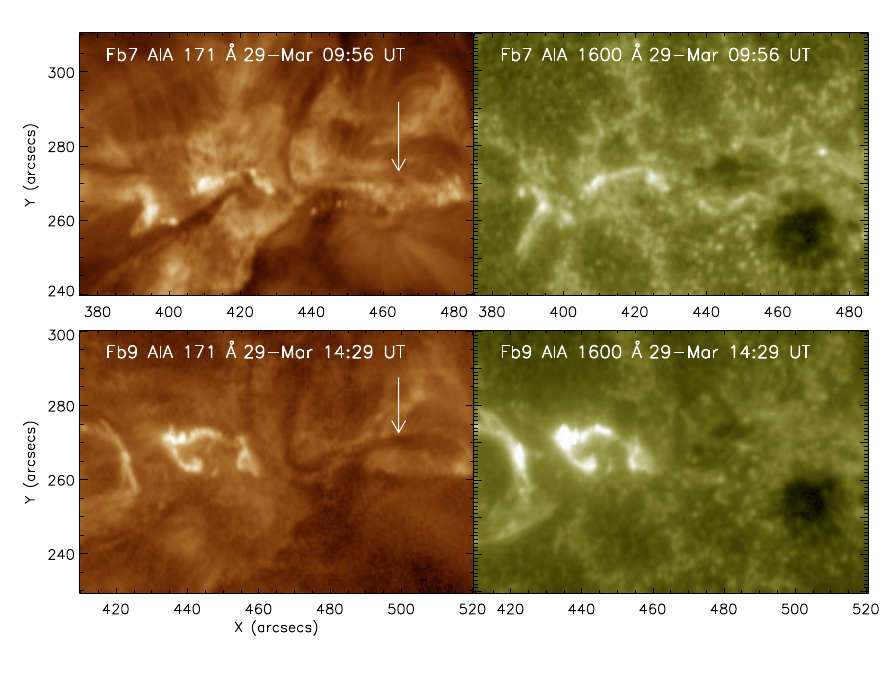} 
\caption{AIA 193 \AA \ and 1600 \AA \ images of the flares Fb7 and Fb9. Note that in the \textit{GOES} soft X-ray flare list, these flares are marked as events occurring in AR 12017. The arrows denote the filament, which we are interested in.}
\label{fig:3}
\end{figure}

The active region 12017 was very flare-productive during the period of 2014 March 28--29 (Figure \ref{fig:1}).
From March 30, there still occurred several more flares in this active region; however, the locations are close to the solar limb, so that the observed magnetic field on the photospheric boundary is not accurate enough for performing an extrapolation for further analysis.
Thus, we only select the flares occurring during March 28--29 as the sample for our study.
As seen from the \textit{GOES} soft X-ray flare list\footnote[3]{\url{ftp://ftp.ngdc.noaa.gov/STP/space-weather/solar-data/solar-features/solar-flares/x-rays/goes/2014/goes-xray-report_2014.txt}}, there are 11 C-class, 1 M-class, and 1 X-class flares that occurred in AR 12017 during these two days.
After checking the observations in 193 \AA \ by the Atmospheric Imaging Assembly (AIA; \citealt{2012Lemen}) on board the \textit{Solar Dynamics Observatory} (\textit{SDO}), we found that most of the flares are likely caused by the eruption of a filament, which keeps existing in the active region for several days (Figure \ref{fig:2}).
However, there are some few exceptions in the flare list.
It is clear that the two C-class flares peaking at 10:00 UT and 14:32 UT on March 29 are not associated with this filament (Figure \ref{fig:3}).
Nevertheless, for the M-class flare peaking at 23:51 UT on March 28, although the \textit{GOES} flare list does not mark it as hosted by the active region, it is clearly caused by the eruption of the filament as revealed by AIA 193 \AA \ observations (Figure \ref{fig:2}). 
Therefore, our sample excludes the two C-class flares but includes the M-class flare as mentioned above.
In total, we have 12 flares in the sample and label them as F1--F12 in the sequence of occurrence time (Table \ref{t:1}).

The AIA 193 \AA \ images (Figure \ref{fig:2}) show that the filament does exist throughout these flares.
However, the filament evolves and undergoes some (partial) eruptions during the relatively long period.
During some flares like F1, it can be seen that the filament contains a right-handed twist.
This implies that the filament can be regarded as, at least part of, an MFR.
Among the twelve selected flares, three of them (F3, F5, and F12) are caused by successful eruptions of the filament and associated with coronal mass ejections (CMEs), which can be detected by \textit{SDO}/AIA and Large Angle and Spectrometric Coronagraph (LASCO).  
Other 	flares are caused by confined eruptions of the filament.

\subsection{Magnetic Field Observations} \label{Sect.2.2}
The magnetic field data come from the Helioseismic and Magnetic Imager (HMI; \citealt{2012Scherrer,2012Schou}) on board \textit{SDO}.
In practice, we use the data from the Space-weather HMI Active Region Patches (SHARPs; \citealt{2014Bobra}), in which the 180\arcdeg \ ambiguity in the horizontal field has been resolved by the minimum energy method \citep{1994Metcalf,2006Metcalf,2009Leka}, the coordinate system of the data has been remapped to a heliographic cylindrical equal-area coordinate system and the vector field has been transformed into the $B_r$, $B_\theta$, and $B_\phi$ components \citep{1990Gary,2013Sun}.

\begin{figure}
\centering
\includegraphics[width=0.72\textwidth]{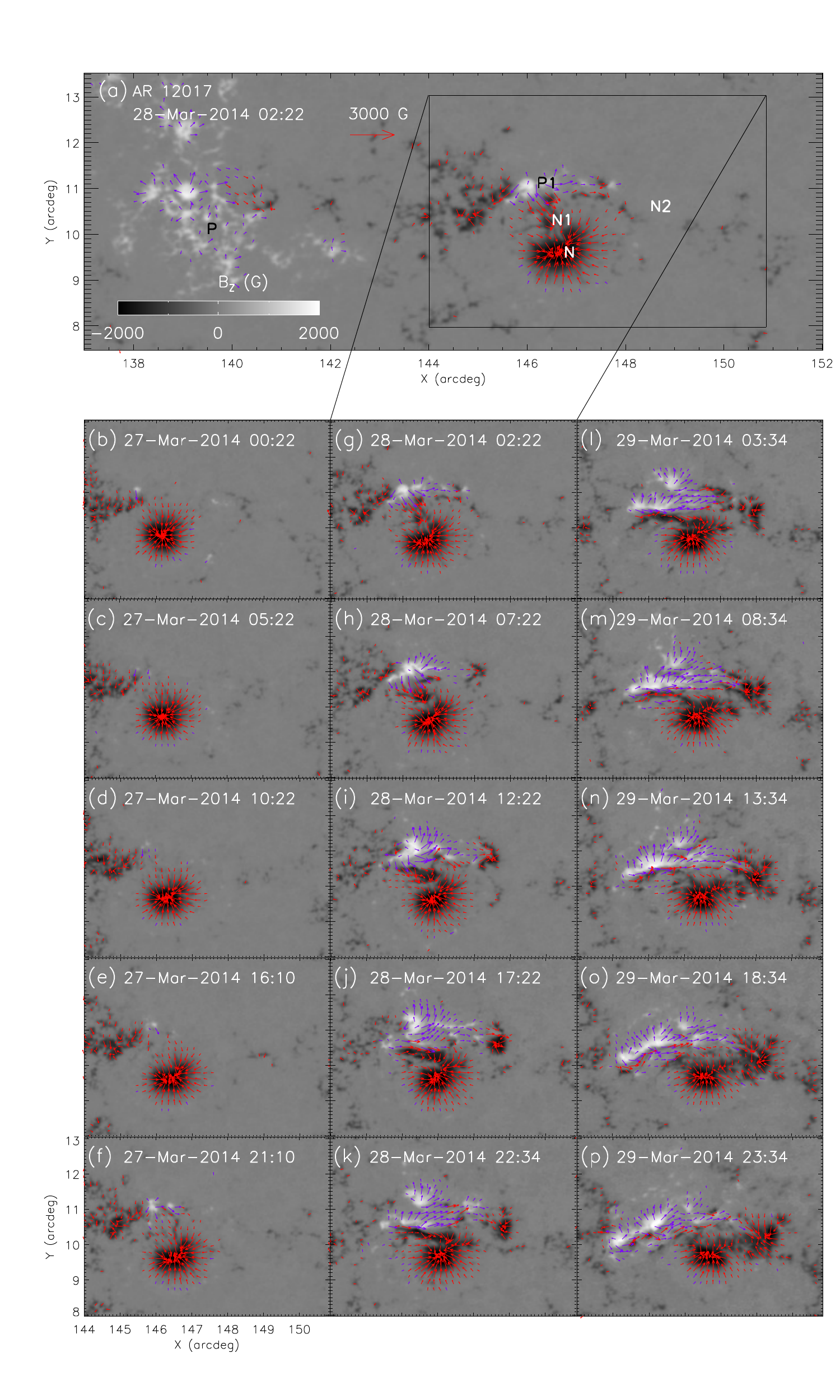} 
\caption{Magnetograms of AR 12017 at selected times from March 27 to 29 observed by HMI. Symbols P and N represent the original positive and negative poles. The three main emerging flux regions are denoted by P1, N1, and N2, respectively. The red and purple arrows indicate the horizontal component of the magnetic field. The heliographic Cylindrical Equal-Area coordinate system is used here, as well as in Figure \ref{fig:5}.}
\label{fig:4}
\end{figure}

We show the time sequence of the magnetic field of AR 12017 from March 27 to 29 in Figure \ref{fig:4}.
AR 12017 had been existing for quite a long time before March 28 when it became flare-productive.
From March 24 to about March 28 12:00 UT, it kept quiet without any flare activities.
During this time interval, the magnetic field was simply a bipole structure.
With time going on, the leading negative pole (N) kept compact while the following positive one (P) became diffused (Figure \ref{fig:4}(a)).
Then, at about 22:00 UT of March 27, a new flux emerged into this active region near the pole N (see Figures \ref{fig:4}(b)--\ref{fig:4}(p)).
This flux emergence makes AR 12017 more active and in particular flare- and CME-productive since about 13:00 UT of March 28.
In more detail, the emerging flux, located to the north of the compact pole N, comprises of three component: a positive pole (P1) and two negative poles (N1 and N2) (Figure \ref{fig:4}(a)).
P1 showed a clear motion toward the east and so did N1; while P2 moved toward the west.

\section{MAGNETIC FIELD ANALYSIS} \label{Sect.3}
\subsection{Magnetic Energy and Helicity Injection} \label{Sect.3.1}
The energy of magnetic field is considered as the dominant energy in the solar corona.
The build-up of the magnetic energy is caused by the Poynting flux transferred from the boundaries.
Here, we only compute the Poynting flux through the photosphere based on the observations.
\begin{equation}\label{eq:e1}
\frac{dE}{dt} = \int_{S} \mathbf{B\times(v\times B)\cdot \hat{n}} d S \text{ ,}
\end{equation}
where $\mathbf{v}$ is the plasma velocity on the photosphere and $\mathbf{\hat{n}}$ is the unit vector normal to the photosphere.

On the other hand, the variation of the relative magnetic helicity of a volume can be written as \citep{1984Berger,2003Demoulin}:
\begin{equation}\label{eq:h1}
\frac{dH}{dt} = -2\int_{S}((\mathbf{v\cdot A}_p)\mathbf{B} - (\mathbf{B\cdot A}_p)\mathbf{v})\cdot \mathbf{\hat{n}}dS\text{ .}
\end{equation}
Here, the subscript ``p" denotes the quantities for a potential field, which is taken as the reference field as usual. 
The helicity flux density can then be defined as $G_{A} = -2((\mathbf{v\cdot A}_p)\mathbf{B} - (\mathbf{B\cdot A}_p)\mathbf{v})\cdot \mathbf{\hat{n}}$ from Equation (\ref{eq:h1}).
However, the flux density defined in this way may produce artifical polarities even without magnetic helicity injection into the corona \citep{2005Pariat}.
Therefore, \cite{2005Pariat} proposed a new formula for calculating the helicity injection rate:
\begin{equation}\label{eq:h2}
\frac{dH}{dt}=-\frac{1}{2\pi}\int_S\int_{S'}\frac{d\theta({\mathbf{r}})}{dt}B_n B_n^{'} dS dS'\text{ ,}
\end{equation}
where
\begin{equation}
\frac{d\theta(\mathbf{r})}{dt}=\frac{1}{r^2}(\mathbf{r}\times\frac{d\mathbf{r}}{dt})_n=\frac{1}{r^2}(\mathbf{r\times(u-u')})_n\text{ ,}
\end{equation}
where $\mathbf{r = x-x'}$ is the relative position of two points ($\mathbf{x}$ and $\mathbf{x}'$) and $d\theta/dt$ represents the corresponding rotation rate.
The new flux density is thus in the form:
\begin{equation}
G_\theta (\mathbf{x}) = -\frac{B_n}{2\pi}\int_{S'}\frac{d \theta(\mathbf{r})}{dt}B_n^{'}d S'\text{ .}
\end{equation}
This flux density can be interpreted as the summation of all other flux elements rotating around one point weighted by their magnetic flux times the normal magnetic component at point $\mathbf{x}$.

In order to calculate the energy and helicity injections through the bottom boundary using Equations (\ref{eq:e1}) and (\ref{eq:h2}), both the vector magnetic field and the vector velocity on the photoshpere are needed.
The vector magnetic field comes from the SHARPs \citep{2014Bobra} and the vector velocity on the photoshpere is derived by applying the differential affine velocity estimator for vector magnetograms (DAVE4VM; \citealt{2008Schuck}) to the SHARPs data base.
The window size used for DAVE4VM is selected as 23 pixels.

\begin{figure}
\centering
\includegraphics[width=1.0\textwidth]{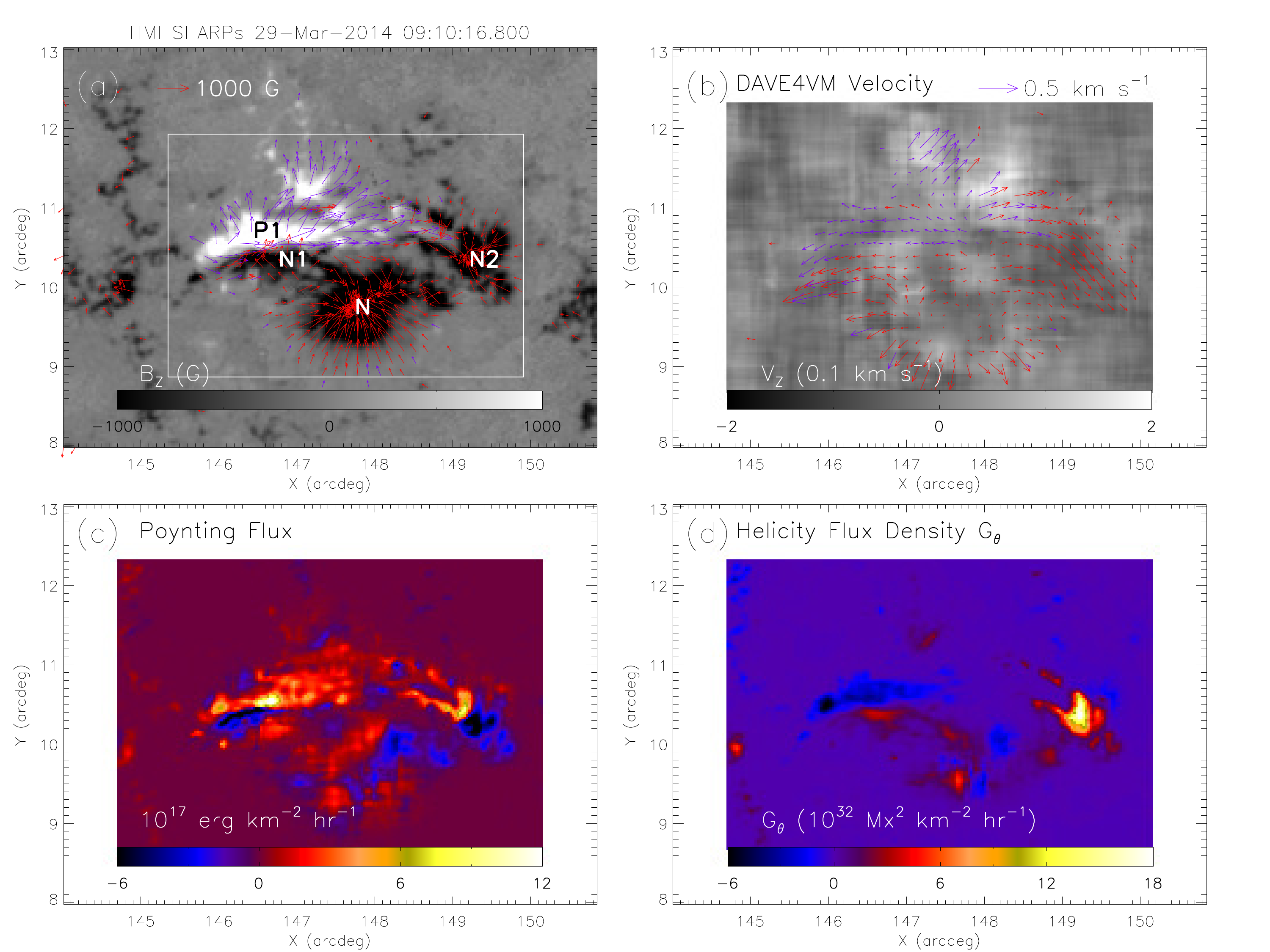} 
\caption{(a) Vector magnetic field from SHARPs data set at 09:10:16 UT on 2014 March 29. Symbol N represents the original negative pole in this active region. Symbols P1, N1, and N2 refer to the three main emerging flux regions, respectively. The white box in panel (a) denotes the field of view of Figures \ref{fig:7}(a) and \ref{fig:10}(a). (b) Vector velocity field derived by the DAVE4VM method at 09:16:16 UT on 2014 March 29. (c) Poynting flux and (d) helicity flux density distributions at the same time as in panel (b).}
\label{fig:5}
\end{figure}

We select an area containing the new emerging flux and compute the energy and helicity injection rate through it.
An example for the selected area and the computed results are shown in Figure \ref{fig:5}.
As mentioned above, the emerging flux is located to the north of the main negative pole N (Figure \ref{fig:5}(a)).
The velocity map confirms that the poles P1 and N1 have an eastward motion and N2 has a westward motion (Figure \ref{fig:5}(b)).
From the Poynting flux map (Figure \ref{fig:5}(c)), it is seen that the energy injection rate associated with P1 is positive, while that associated with N1 is negative.
N2 seems to contribute both positive and negative energy fluxes (Figure \ref{fig:5}(c)).
On the other hand, it is found from the helicity injection map (Figure \ref{fig:5}(d)) that P1 emerges with negative helicity injection while N1 and N2 emerge with positive helicity injection into the corona (Figure \ref{fig:5}(d)).

\begin{figure}
\centering
\includegraphics[width=0.79\textwidth]{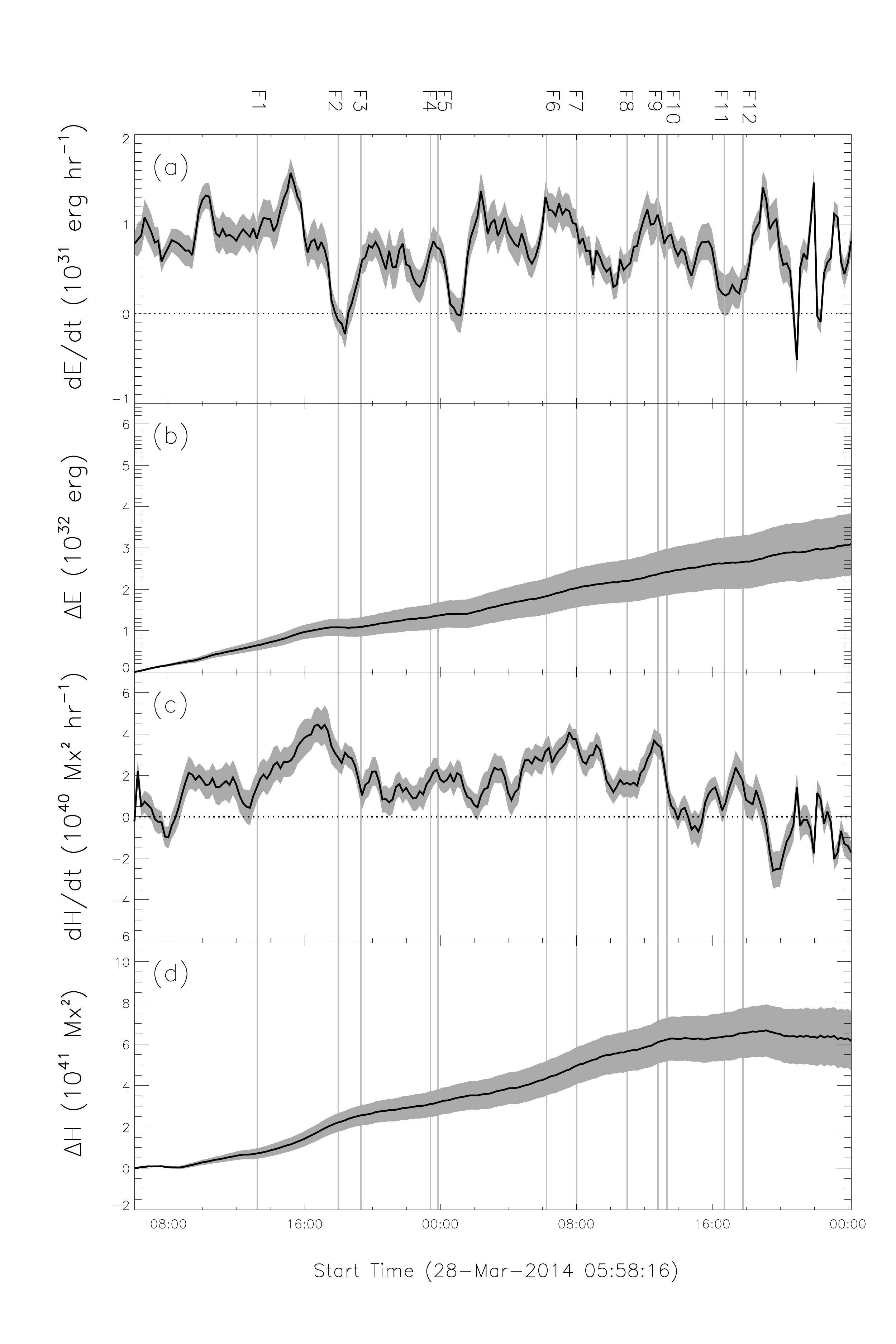} 
\caption{(a) Evolution of the spatially integrated energy injection rate . (b) Time integrated magnetic energy, $\Delta E = \int_0^{\Delta T} (dE/dt) \text{ }dt$, where $\Delta T$ donates the time interval from 23:58 UT of March 27 to the specific time for calculation. (c) Injection rate of relative magnetic helicity from the new emerging flux. (d) Time integrated relative magnetic helicity, $\Delta H = \int_0^{\Delta T} (dH/dt)\text{ }dt$, where $\Delta T$ has the same meaning as in (b). The shaded areas indicate the standard deviation of each quantities from ten times calculation. The peak time of each flare is shown by a vertical line.}
\label{fig:6}
\end{figure}

Since the vector magnetic field has a measurement error, we artificially add a noise (a random coefficient times the measurement error from the SHARPs error file) to the observed data, and then compute the vector velocity and the associated energy and helicity injection for 10 times.
In Figure \ref{fig:6}, we show the mean value of the energy and helicity injection rates with the solid curves and the standard deviations by the shaded areas.
The vertical lines represent the peak time of the flares as in Figures \ref{fig:6}, \ref{fig:8}, \ref{fig:11}, \ref{fig:12}, and \ref{fig:13}.
We find that the spatially integrated Poynting flux and helicity injection rates are mostly positive and in the order of $10^{31}$ erg hr$^{-1}$ and $10^{40}$ Mx$^2$ hr$^{-1}$, respectively (Figures \ref{fig:6}(a) and \ref{fig:6}(c)).
They appear to be negative only in a few cases and with short time intervals.
A further temporal integration of the Poynting flux and helicity injection flux with time yield the accumulated energy and helicity in the volume of interest through the bottom boundary (Figures \ref{fig:6}(b) and \ref{fig:6}(d)).
It is seen that the accumulated energy and helicity are nearly monotonically increasing.

\subsection{Magnetic Field Extrapolation} \label{Sect.3.2}
To obtain the 3D magnetic field of the corona from March 28 to 29, we adopt the nonlinear force-free field extrapolation with the optimization method \citep{2000Wheatland,2004Wiegelmann} using the SHARPs data (with a cadence of 12 minutes) as the prescribed condition at the bottom boundary.
Although these data have already undergone a series of preprocessing to resolve the 180\arcdeg \ ambiguity and the projection effect \citep{2013Sun,2014Bobra}, the field may deviate more or less from the force-free and torque-free state.
Therefore, an additional preprocessing is applied to remove the net force and torque on the boundary before the extrapolation \citep{2006Wiegelmann}.

Based on the extrapolated 3D field, we can successfully identify an MFR (see Section \ref{Sect.3.3}).
In Figures \ref{fig:7}(a), \ref{fig:7}(c), and \ref{fig:7}(d), we plot some selected field lines for the MFR and the ambient field at 09:10:16 UT on March 29.
From a comparison with the AIA 304 \AA \ and 171 \AA \ images (Figure \ref{fig:7}), the MFR is shown to be cospatial with the filament whose eruption may drive the occurrence of the flares (CMEs) in this active region during the two days.
To calculate the magnetic energy of the MFR, we select a volume encompassing the MFR, which is shown as the white box in Figure \ref{fig:7}(a).
For estimating the uncertainty in energy calculation, we have repeated the extrapolations 10 times with a random noise (as described above) added to the original boundary data.
The standard deviation is shown as the shaded area around the mean value in Figure \ref{fig:8}.
The magnetic energy of the potential field and that of the NLFFF contained in this box are shown in Figure \ref{fig:8}(a).
It is seen that the potential energy changes little, while the NLFFF energy varies much more significantly with time.
The free magnetic energy, defined as the difference between the NLFFF energy and the potential energy, is shown to be in the order of $10^{31}$ erg (Figure \ref{fig:8}(b)).
The occurrence of the X-class flare (F12) is accompanied by a decrease of the free energy of $1\times10^{31}$ erg.
This decrease is slightly smaller than that estimated by \cite{2015Aschwanden} ($(29\pm3)\times10^{30}$ erg).
The difference might be caused by a systematic error from different methods and even the inconsistency in the bottom boundary condition after the preprocessing of the data as performed above.

The force-freeness of the extrapolations is checked with the metric used by \cite{2000Wheatland}, current weighted sine of the angle between the current field ($\mathbf{J} = \frac{1}{\mu_0}\nabla\times\mathbf{B}$) and the magnetic field ($\mathbf{B}$):
\begin{equation}
\langle\textrm{CW} \textrm{sin}\theta\rangle = \frac{\sum_{i}J_i \textrm{sin}\theta_i}{\sum_{i}J_i}\text{ ,}
\end{equation}
where
\begin{equation}
\textrm{sin}\theta_i = \frac{|\mathbf{J}_i\times\mathbf{B}_i|}{J_i B_i}\text{ .}
\end{equation}
Generally speaking, the average angle defined as $\bar{\theta} = \textrm{arcsin} \langle\textrm{CW} \textrm{sin}\theta\rangle $ in the whole computational domain and in the small box containing the flux rope (shown in Figure \ref{fig:7}(c) and \ref{fig:7}(d)) are in the range of 25\arcdeg -- 35\arcdeg \ and  6\arcdeg -- 20\arcdeg , respectively.
These values are reasonable and similar to those reported in previous papers (e.g., \citealt{2008Schrijver,2010Guo,2012Sun,2015Yang}).
The deviation from force-freeness mainly comes from the fact that the side and top boundaries, on which the potential field is used, are inconsistent with the bottom one, on which we use the observed vector magnetic field after preprocessing.
Next, we check the divergence state of the magnetic field in the extrapolation results.
The dimensionless metric $f_i$ \citep{2000Wheatland} is selected to estimate the divergence-freeness as,
\begin{equation}
f_i=\frac{\int_{\Delta S_i}\mathbf{B}\cdot d\mathbf{S}}{\int_{\Delta S_i}|\mathbf{B}|dS}.
\end{equation}
The average of the absolute value of $f_i$, namely $\langle |f_i| \rangle$ in the whole domain, is in the range of 1.3--1.8$\times10^{-3}$, which is similar to \cite{2010Guo}.

\subsection{QSLs of the flux rope} \label{Sect.3.3}
QSL is a place with very large gradient of the field line connectivity \citep{1995Priest,a1996Demoulin,b1996Demoulin,1997Demoulin,b2006Demoulin,1997Mandrini,2002Titov}, which is usually measured by the squashing factor, $Q$ \citep{2002Titov}.
Since field line connectivities of the MFR and the ambient field are quite different from each other, the value of $Q$ can be very large at the interface between the MFR and the ambient field.
In principle, a QSL may appear around the MFR separating it from the ambient field.
This QSL can be used to determine the geometry of the MFR.

We compute the $Q$ map on a fixed slice for each extrapolation based on the SHARPs data during March 28 to 29.
To show the typical evolutions of the QSL structure, we plot the $Q$ maps around the peak time of eight flares (F2, F3, F4, F5, F6, F9, F10, and F12) in Figure \ref{fig:9}. 
Each column is for one flare, except the columns \ref{fig:9}b and \ref{fig:9}e, each of which includes two flares (F4 and F5, F9 and F10, respectively) that occur closely.
Note that in Figure \ref{fig:9}, the first four rows show the QSL structure before the flares while the last row shows that after the flares (except that panels b4, b5, e4, and e5 refer to the structures just after flares F4, F5, F9, and F10, respectively).
One can find that before the flare the QSL structure either gets larger slowly (columns \ref{fig:9}c and \ref{fig:9}e) or changes in shape slightly (columns \ref{fig:9}a, \ref{fig:9}b, \ref{fig:9}d, and \ref{fig:9}f).
However, if comparing the QSL structure before and after the flare, one can see that it decreases a lot in a short time.
The decrease of the QSL structure means that the cross section of the MFR gets smaller but not disappear, which is likely a consequence of the partial eruption of the flux rope, as has often been observed \citep{2012Liu,2012Shen,2015Zhang}.
We have also checked the QSL structure during the flares F1, F7, F8, and F11, which, however, does not show an abrupt variation as revealed in the former flares.

\begin{figure}
\centering
\includegraphics[width=1\textwidth]{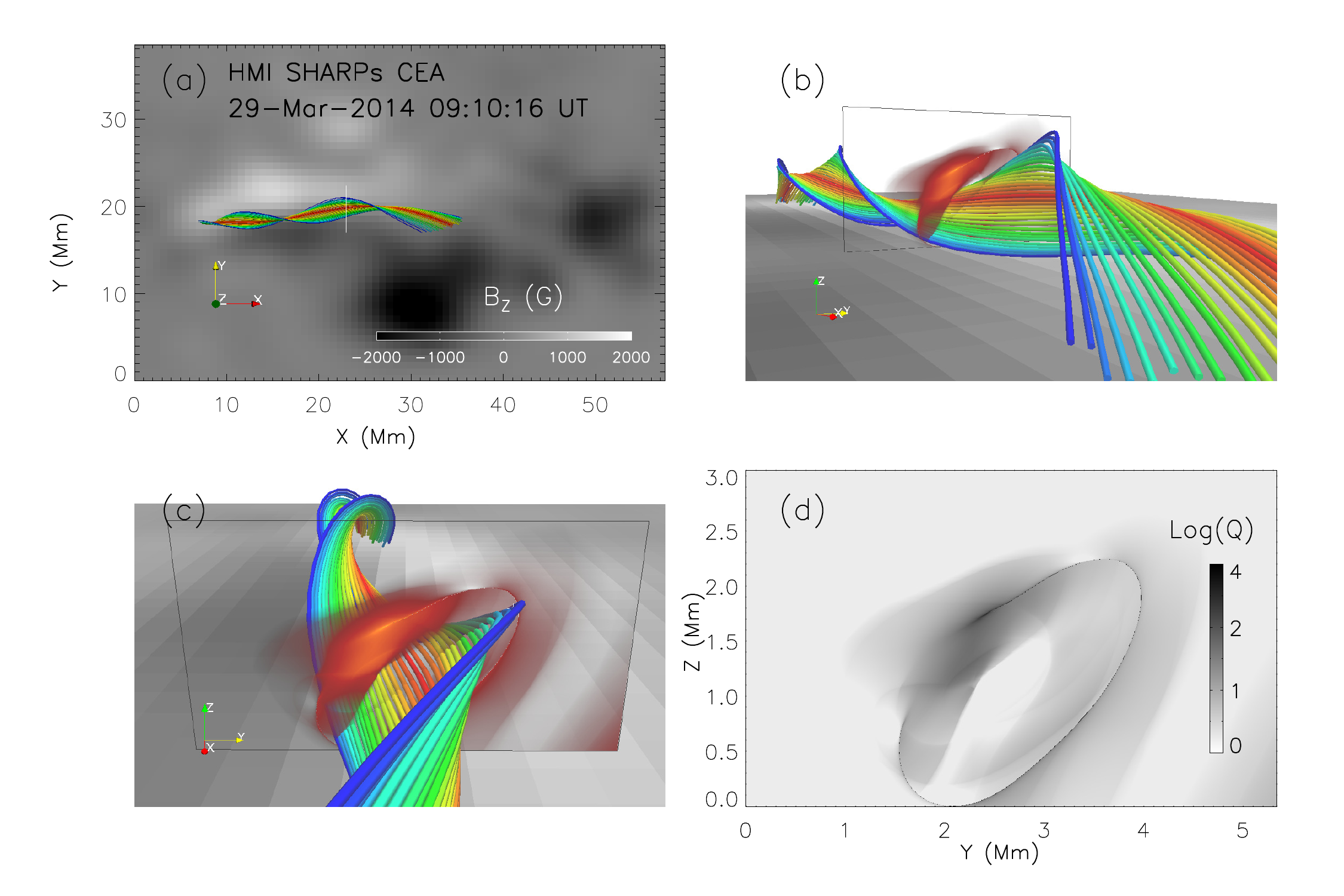} 
\caption{(a) Top view of the MFR from the extrapolation at 09:10:16 UT  on 2014 March 29. The background is the vertical magnetic field, $B_z$, from SHARPs data set. (b)--(c) Side views of the MFR. The black frame indicates the fixed slice at which the squashing factor, $Q$, is calculated. (d) Map of the squashing factor on the fixed slice. }
\label{fig:10}
\end{figure}

From the above analysis, it is clear that the QSL structure shown in the $Q$ maps provides a good definition of the MFR.
In practice, we use the following steps to determine the field lines threaded within the MFR.
First, we use the method proposed by \cite{2012Pariat} to calculate the value of the squashing factor, $Q$, on a fixed slice cut through the middle of the MFR.
This $Q$ map can give a closed QSL structure around the MFR.
Second, the closed QSL indicates the outer surface of the MFR, we select the sample locations uniformly inside this outer surface.
Third, we integrate the field lines running through the selected locations as those composing the MFR.
As an example, we show in Figure \ref{fig:10} the field lines of an MFR at 09:10:16 UT of March 29, whose mapping on the slice is clearly within the closed QSL structure.
Generally, these field lines can well define the topology and magnetic structure of an MFR.

\subsection{Decay index} \label{Sect.3.6}
As the MFR has been well defined, we calculate the decay index \citep{2006Kliem} of the ambient field:
\begin{equation}
n = -\frac{h}{B_p}\frac{\partial B_{p}}{\partial h},
\end{equation}
where $B_{p}$ is the horizontal potential field.
The decay index is shown as contours overplotted on the $Q$ maps (Figure \ref{fig:9}).
It is seen that the value of the decay index around the closed QSLs varies in a range of $0.1$--$0.6$.
We also check the temporal variation of this parameter throughout the two days and find the highest value to be around $0.6$, which is less than the critical value for the torus instability \citep{2006Kliem,2010Demoulin,2015Myers}.
This implies that the MFR could be stable under the balance between its hoop force and the constraining force of the overlying field.

\subsection{Twist number estimation}\label{Sect.3.4}
Using the field lines consisting of the MFR, which are obtained with the method described in Section \ref{Sect.3.3}, we estimate the twist number of the MFR.
For a further comparison, two methods are used here.
The first method is the widely used parallel current integration along the field line \citep{2006Berger}:
\begin{equation}\label{eq:t1}
T_{1} = \frac{1}{4\pi}\int_s\frac{\nabla\times\mathbf{B}(s)\cdot\mathbf{B}(s)}{|\mathbf{B}(s)|^2}ds,
\end{equation}
where $s$ is the length parameter of the field line. 
We then define the associated average twist number as:
\begin{equation}\label{eq:taver}
\langle T_{1}\rangle=\frac{\sum_{i}\Phi_i\text{ } T_{1,i}}{\sum_{i}\Phi_i},
\end{equation}
where $\Phi_i$ is the magnetic flux of flux tube $i$.
The second one is to calculate the rotation of a secondary curve around the axis \citep{2006Berger,2013Guo}:
\begin{equation}\label{eq:t2}
T_{2} = \frac{1}{2\pi} \int_{s} \mathbf{T}(s)\cdot\mathbf{V}(s)\times\frac{d\mathbf{V}(s)}{d s} d s,
\end{equation}
where $\mathbf{T}$ denotes the unit tangent vector of the axis curve and $\mathbf{V}$ is the unit vector normal to $\mathbf{T}$ and pointing from the axis curve to the secondary curve.
We calculate this twist number ($T_{2}$) by choosing one field line as the axis curve and the other one as the secondary curve, and then repeat the calculation by exchanging the role of these two curves.
The average twist number calculated with the second method is labelled as $\langle T_{2}\rangle$:
\begin{equation}
\langle T_{2}\rangle = \frac{\sum_{i,j}\Phi_i\Phi_j \text{ } T_{2,ij}}{\sum_{i,j}\Phi_i\Phi_j}.
\end{equation}

\begin{figure}
\centering
\includegraphics[width=0.84\textwidth]{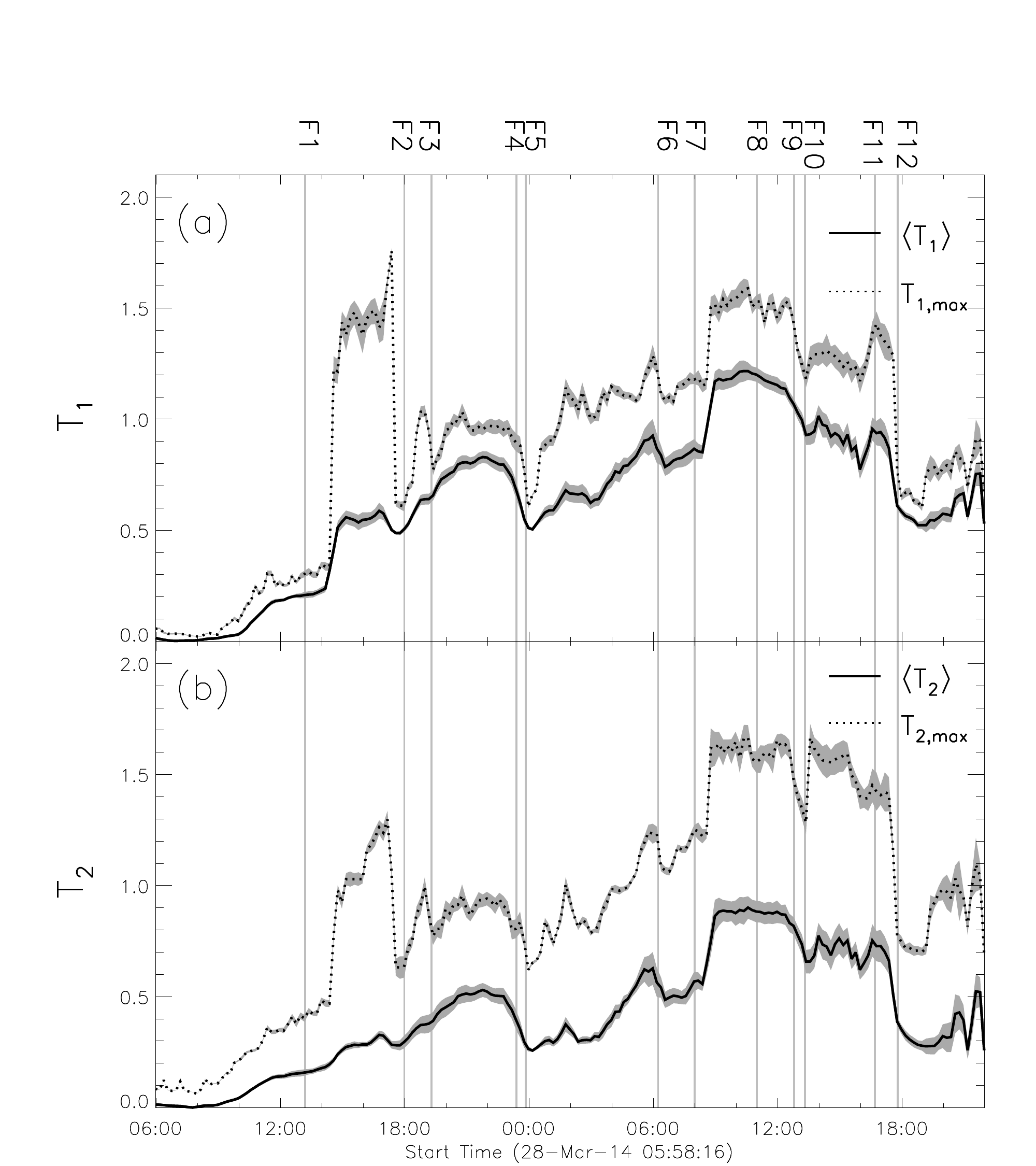} 
\caption{Time profiles of the twist numbers. The solid and dotted lines represent the average twist numbers and the
corresponding maximum twist numbers of the MFR, respectively. (a) The twist is computed by integrating the torsional
parameter $\alpha$ along the field lines. (b) The twist is computed by the rotation rate of two field lines. The vertical lines and the shaded areas have the same meaning as in Figure \ref{fig:6}.}
\label{fig:11}
\end{figure}

To estimate the uncertainty in the twist number, we change the sample locations inside the QSL structures and repeat the calculation of the twist number for 10 times.
The mean values and the standard deviations are shown by the solid curves and the shaded areas in Figure \ref{fig:11}.
Note that the vertical lines have the same meaning as in Figure \ref{fig:6}.
The value of $\langle T_{1}\rangle$ (Figure \ref{fig:11}(a)) is a slightly larger than that of $\langle T_{2}\rangle$ (Figure \ref{fig:11}(b)), reflecting the intrinsic difference of the two methods.
However, their time profiles are almost the same.

In addition, the twist number of the MFR is not distributed uniformly. 
Within the cross section of the MFR, the maximum value of the twist number calculated by Equations \ref{eq:t1} and \ref{eq:t2} is referred to as the maximum twist number ($T_{1,max}$ and $T_{2,max}$) hereafter.
Their errors are estimated similarly to that of the average twist number. 
It is found that the maximum twist number is obviously larger than the average one (Figure \ref{fig:11}). 
We further calculate quantitatively the twist number decrease at the flare peak time with respect to that half an hour before.
The result is summarized in Table \ref{t:2}. 
It can be seen from Table \ref{t:2} and Figure \ref{fig:11} that some flares (e.g., F5 and F12) occur in association with a sharp decrease of the twist number, especially the maximum twist number.
At the initial stage (12:00 UT to 20:00 UT on March 28), the time profiles of the maximum twist number are somewhat different from that of the average one.
A decrease by as large as $65\%$($40\%$) of the maximum twist number $T_{1,max}$($T_{2,max}$) appears before F2, while the average twist number $\langle T_{1}\rangle$($\langle T_{2}\rangle$) shows only a $1\%$($4\%$) variation.
However, the maximum twist number $T_{1,max}$($T_{2,max}$) decreases by $24\%$($22\%$), but the average twist number $\langle T_{1}\rangle$($\langle T_{2}\rangle$) increases by $3\%$($5\%$) before F3.
This suggests that the MFR consists of different parts having quite different properties from the average property of the MFR as a whole.

On the other hand, we find six flares that are associated with a sharp decrease by more than 10\% of the maximum twist number (F2, F3, F5, F9, F10, and F12), a threshold obviously higher than the uncertainty in the calculations (Table \ref{t:2}).
Among them, F2 and F12 show a decrease by about half in the maximum twist number, and the decrease for the other four flares range from about $10\%$ to $30\%$.
Note that flares F4 and F6 also show a clear decrease in the maximum twist number, though it is relatively small (Figure \ref{fig:11}).

\subsection{Magnetic helicity of the flux rope} \label{Sect.3.5}
Magnetic helicity is a conserved physical quantity in ideal MHD.
In a closed volume without field lines running in or out of the surface, the magnetic helicity is usually written as $H^{closed} = \int_{V}\mathbf{A\cdot B}d\mathcal{V}$.
Under the Coulomb gauge, it can be reformulated as \citep{a2006Demoulin}:
\begin{equation}\label{eq:h}
H^{closed} = \int_{\Phi}\int_{\Phi'}\mathcal{L}_{\mathcal{C,C'}}^{closed} d\Phi d\Phi'\text{ ,}
\end{equation}
where
\begin{equation}
\mathcal{L}_{\mathcal{C,C'}}^{closed} = \frac{1}{4\pi}\oint_{\mathcal{C}}\oint_{\mathcal{C'}} \frac{\mathbf{B}}{|B|}\times\frac{\mathbf{B'}}{|B'|}\cdot \frac{\mathbf{x}-\mathbf{x'}}{|\mathbf{x}-\mathbf{x'}|^3}dldl'\text{ .}
\end{equation}
This integral is taken on two closed magnetic field lines $\mathcal{C}$ and $\mathcal{C'}$, whose corresponding magnetic fluxes are $\Phi$ and $\Phi'$, respectively.
The quantity $\mathcal{L}_{\mathcal{C,C'}}^{closed}$ is known as the Gauss linking number \citep{2006Berger,a2006Demoulin}, which gives the number of times that one flux tube winds around the other.
Therefore, the magnetic helicity of two closed flux tubes can be interpreted as the winding number times the magnetic fluxes of each flux tubes, and the total magnetic helicity in the volume refers to the summation of the helicity of each pair of the flux tubes \citep{a2006Demoulin}.
A further derivation given by the Equation (37) in \cite{a2006Demoulin} shows that the ratio of mutual to self helicity increases with flux tube number; when the flux tube number is large enough, the self helicity could be neglected.

However, Equation (\ref{eq:h}) cannot be applied to a practical case, which is usually an open magnetic configuration like the magnetic field in solar corona.
In such a case, the relative magnetic helicity, $H_R = \int_{V} (\mathbf{A}+\mathbf{A}_R)\cdot(\mathbf{B}-\mathbf{B}_R)\rm d \mathcal{V}$, should be applied \citep{1984Berger}.
Recently, some numerical methods for computing the relative magnetic helicity of active regions have been developed \citep{2011Rudenko,2011Thalmann,2012Valori,2013Yang}, which, however, need a reference field and the vector potential.
In particular, \cite{a2006Demoulin} proposed a method to calculate the relative magnetic helicity.
The main idea of their method is to calculate the mutual helicity of each pair of discrete flux tubes, which consist of the MFR, and then sum all of them, $H^{mutual} = \sum_{i,j} \mathcal{L}_{i,j}\Phi_i\Phi_j$ \citep{a2006Demoulin}.
For two tubes $i$ and $j$, $\Phi_i$ and $\Phi_j$ indicate the magnetic fluxes and the parameter $\mathcal{L}_{i,j}$ is the corresponding mutual helicity, which can be decomposed into two parts, $\mathcal{L}_{i,j} = \mathcal{L}_{i,j}^{closed}+\mathcal{L}_{i,j}^{arch}$.
Here, the quantity $\mathcal{L}_{i,j}^{closed}$ is the Gauss linking number and $\mathcal{L}_{i,j}^{arch}$ is a mutual helicity that is defined by the relative foot-point position.
Detailed descriptions of the parameter $\mathcal{L}_{i,j}^{arch}$ can be found in \cite{a2006Demoulin} and \cite{2012Georgoulis}.
This method has an advantage that it avoids calculating the reference field and the vector potential. 
Moreover, it can be used to compute the relative magnetic helicity for one isolate topological system, whose shape is not regular, with enough field lines threaded within this system, e.g., an MFR.

We adopt the method of \citep{a2006Demoulin} to compute the mutual helicity, $H^{mutual}$, of the MFR, whose field lines have been obtained by the method described in Section \ref{Sect.3.3}.
Since we have decomposed the MFR into thousands of winding flux tubes, the mutual helicity is about three orders of magnitude larger than the self helicity, as has been predicted by \cite{a2006Demoulin}.
Thus, the mutual helicity can represent the total relative magnetic helicity of such a system (Figure \ref{fig:12}(d)).
While the above method is relatively strict, we can also use the relationship between the helicity and the twist number to estimate the helicity ($H=T\Phi^2$) based on the two kinds of average twist numbers ($\langle T_{1}\rangle$ and $\langle T_{2}\rangle$) calculated in Section \ref{Sect.3.4} , which are labelled as $H_{twist,1}$ and $H_{twist,2}$ (Figures \ref{fig:12}(b) and \ref{fig:12}(c)).
Thus, we get three values of magnetic helicity estimations that are shown to be quantitatively different; more specifically, the values of $H_{twist,1}$ and $H_{twist,2}$ are about 1.7 and 1.3 times the value of $H^{mutual}$, respectively.
Such a difference is not unexpected considering the quite different methods used.
The method for $H_{twist,1}$ and $H_{twist,2}$ is an approximate one that depends on the prior calculation of parameters $\langle T_{1}\rangle$, $\langle T_{2}\rangle$, and $\Phi$, while $H^{mutual}$ is calculated from a direct integration that suffers from less errors.
Nevertheless, it is clear that the different methods yield quite similar time profiles of the helicity, which suggests that the results are reasonable and convincing.
It is interesting that the magnetic helicity evolves in a different manner from the twist number (including quantities $\langle T_{1}\rangle$, $\langle T_{2}\rangle$, $T_{1,max}$, and $T_{2,max}$).
The results show that the magnetic helicity does not change too much on March 28 and in particular varies differently from the twist number during the flaring period (Figure \ref{fig:11}).
For example, in the C-class flare F8, the magnetic helicity shows an increase by approximately $50\%$ in half an hour before the flare peak but much smaller changes are detected for the maximum and average twist numbers (Table \ref{t:2}).
A somewhat different behavior is found for the X-class flare F12, in which the magnetic helicity shows a slight decrease.
Such a change is more likely the consequence of a long-term evolution rather than a response to the flare.
By contrast, the twist number has a sharp decrease during the flare, as mentioned in Section \ref{Sect.3.4}.
On the other hand, if we compare the mutual helicity ($H^{mutual}$) with the time integral helicity through the emerging flux (Figure \ref{fig:13}(b)), the former amounts to only several percentages of the latter, confirming the previous finding by \cite{2013Guo}.

\section{DISCUSSION AND SUMMARY} \label{Sect.4}
\subsection{Variation of QSLs}
The closed QSL surrounding the MFR provides a good delineation of the latter in two dimensions (the cross section).
We show the 2D $Q$-maps at a number of time instants around eight flares (Figure \ref{fig:9}).
A general behavior regarding the evolution of the QSL structure can be found from them. 
Before the flare, the QSL structure varies slowly; however, it decreases sharply in cross section in a short time during the occurrence of a flare.
The decrease of the cross section of the QSL structure suggests that the MFR may be partly erupted during the flare \citep{2006Gibson}.
The remainder of the MFR can evolve (grow larger in shape) gradually and drive another flare with the continuous energy and helicity injection into the corona (Section \ref{Sect.3.1}).
Note that the QSL does not show a significant variation during other small flares, neither do other parameters like the energy and helicity injections, the free magnetic energy, the twist number, and the magnetic helicity (Figures \ref{fig:6}, \ref{fig:8}, \ref{fig:11}, and \ref{fig:12}).

\subsection{Trigger mechanism of the MFR}
To investigate what factors are responsible for the MFR eruption, we calculate the decay index of the ambient potential field \citep{2006Kliem}.
The decay index at the site of the MFR is found to be in the range of $0.1$--$0.6$, which is less than the critical value for the torus instability \citep{2006Kliem,2010Demoulin,2015Myers}.
This indicates that the torus instability is unlikely the main factor causing the MFR eruption.

On the other hand, we explore the possible role of the kink instability, for which the twist number is the key parameter \citep{2003Fan,2005Fan,2005Torok}.
Theoretically, the kink instability occurs when the twist number exceeds some critical value.
For this end, we calculate the twist number of the MFR that is plotted in Figure \ref{fig:11}.
However, from the observational aspect, we do not find a common critical value of the twist number over which the MFR tends to erupt leading to the occurrence of a flare and/or a CME.
A possible reason is that the critical value may depend on the specific geometry and other physical conditions of the MFR, like the loop aspect ratio, plasma beta, transverse field, and line-tying effect \citep{1979Hood,1981Hood,2004Torok}.
Since the MFR evolves with time, these parameters all change and so does the critical value of twist number.
Nevertheless, we can still reach some conclusions from the evolution of the twist number.
First, we find that the maximum twist number ($T_{1,max}$ and $T_{2,max}$) is clearly larger than the average one ($\langle T_{1}\rangle$ and $\langle T_{2}\rangle$).
This implies that some small part of the MFR may contain a locally larger twist number than the MFR as a whole and is likely to trigger a local kink instability.
Second, reductions to some extent of the twist number can be identified in association with the occurrence of some flares.
In particular, the maximum twist numbers ($T_{1,max}$ and $T_{2,max}$) are shown to vary more sharply than the average one ($\langle T_{1}\rangle$ and $\langle T_{2}\rangle$) during flares.

Based on a quantitative evolution of the decay index and the twist number as well as a detailed comparison with observations, we conclude that the flares occurring in AR 12017 during March 28 to 29 are more likely triggered by the kink instability rather than the torus instability of the MFR.
The MFR undergoes a sequence of partial eruptions leading to these flares while the main body of the MFR keeps existing during the whole time period.
This is supported by the fact that the QSL structure remains to exist after each flare though it may change somewhat in shape or size.
Although the MFR has an average twist number that is not large enough, some local part may have a larger twist number that exceeds the critical value for kink instability.
An illustrative example is the flare F3, during which the maximum twist number evolves differently from that of the average one (Figure \ref{fig:11}).
On the other hand, the AIA images during most of the flares (Figure \ref{fig:2}) show clearly that the MFR eruptions are always confined.
This is consistent with the scenario that the decay index of the ambient field is too small to bring about the torus instability.
However, for the three successful eruptions, a possible explanation is that an initial kink perturbation makes the MFR rise to a higher altitude where the torus instability could occur, which is similar to the result of \cite{2016LiuR}.

\subsection{Helicity injection and helicity of the MFR}
We also calculate the energy and helicity injection rates from the bottom boundary, which are shown to be almost positive and not to clearly correlate with the occurrence of flares (Figures \ref{fig:6}(a) and \ref{fig:6}(c)).
The time integrated energy and helicity increase nearly monotonically with no abrupt variation around the flaring time, which is similar to previous findings \citep{2003Sakurai,2008Park}.
Quantitatively, the helicity injection rate is in the order of $10^{40}$ Mx$^2$ hr$^{-1}$.
Thus, a total amount of helicity in the order of $10^{41}$ Mx$^2$ can be accumulated in this active region within several hours before the flare, which is needed for eruptions \citep{1994LowBC,1995Kusano}.
It seems that the energy and helicity injections from the bottom boundary help build an essential physical condition for flare occurrence in the long run.
However, they usually do not play the role of triggering and thus have no instantaneous change in response to single flares.
As have been revealed in some recent studies \citep{2005Torok,2010Aulanier,2014Amari}, the central engine for eruptions is very likely the MFR or the highly sheared field, which in quantity constitutes only a small part of the whole active region in terms of energy and helicity.
Specifically, the twist number of the MFR and the QSL structure surrounding the MFR are shown to change more evidently with the occurrence of flares (Sections \ref{Sect.3.3} and \ref{Sect.3.4}).

\begin{figure}
\centering
\includegraphics[width=0.84\textwidth]{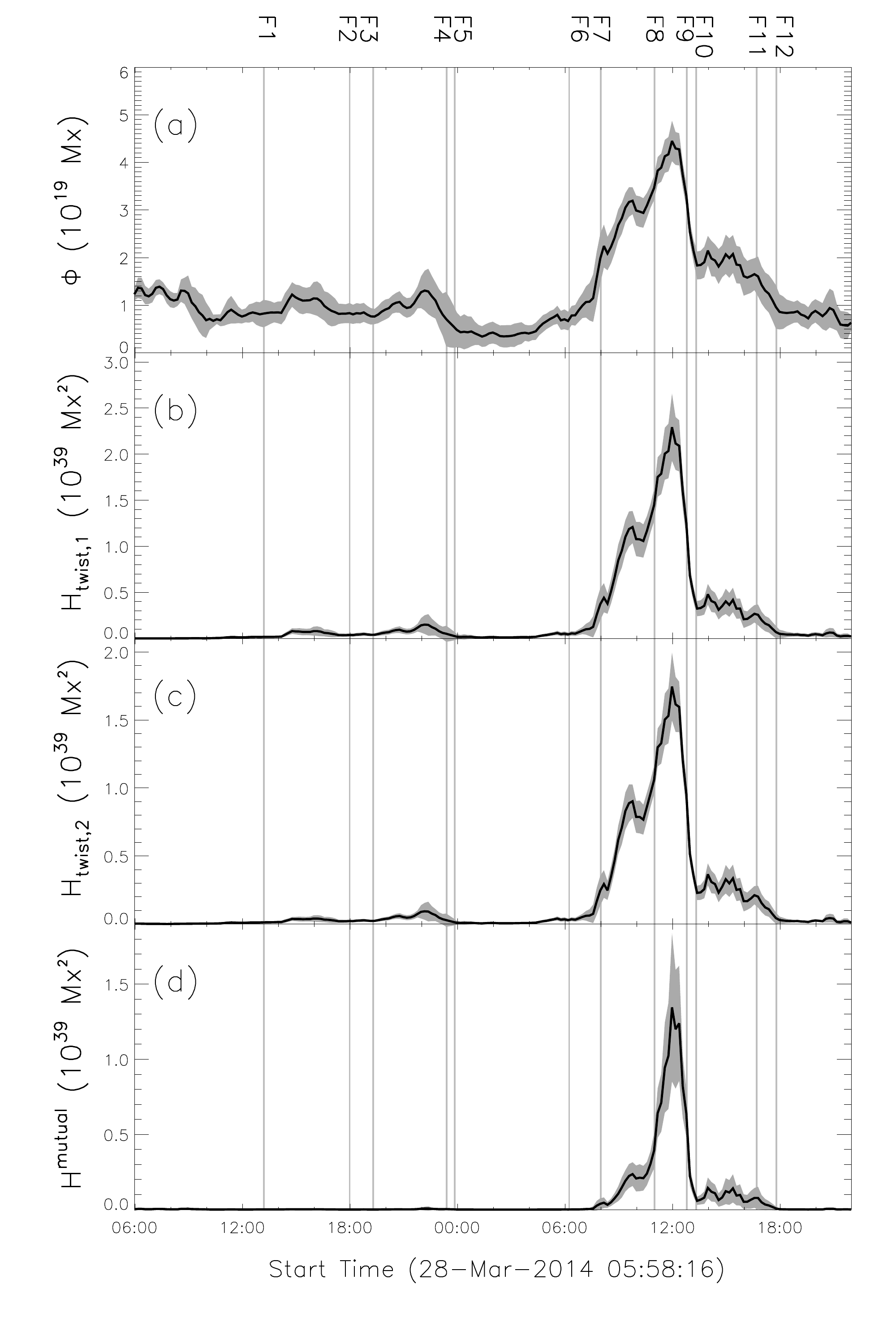} 
\caption{(a) Total magnetic flux within the MFR surrounded by large $Q$ values. (b)--(c) Magnetic helicity, $H_{twist,1}$ and $H_{twist,2}$, and (d) Mutual helicity of the MFR, $H^{mutual}$. The vertical lines and the shaded areas have the same meaning as in Figure \ref{fig:6}.}
\label{fig:12}
\end{figure}

To evaluate the magnetic helicity of the MFR, we have used different methods.
The results show similar time profiles, though quantitatively different (Figures \ref{fig:12}(b), \ref{fig:12}(c) and \ref{fig:12}(d)).
It is found that the quantities, $H_{twist,1}$ and $H_{twist,2}$ are about 1.7 and 1.3 times the value of $H^{mutual}$.
In fact the corresponding estimation for $\langle T_{1}\rangle$ is based on the parallel current integration, while $\langle T_{2}\rangle$ is based on the twist number of two field lines.
The definition of the latter is closer to that of the mutual helicity, both of which describe how much two field lines wind around each other.
As a consequence, the quantity $H_{twist,2}$ is closer to the quantity $H^{mutual}$ in magnitude.
Moreover, comparing the helicity and the twist number, thought closely related, show different behaviors in relation to the occurrence of flares.
The reason is that the magnetic helicity relies not only on the twist number but also on the square of the total magnetic flux.
Sometimes the latter is even more weighted than the former.
For example, the total magnetic flux and helicity of the MFR show an almost synchronous increase from about 6:00 UT to 12:00 UT of March 29 and reach their maximum simultaneously (Figure \ref{fig:12}).
It is worth noting that among these twelve flares, the helicity shows a significant change only in one flare (F9).
By contrast, the twist number, especially the maximum twist number ($T_{1,max}$ and $T_{2,max}$), seems to vary more sensitively to the occurrence of flares.
A possible reason is that the magnetic helicity is a measure of the field line wrapping of the whole MFR; while the maximum twist number is thought to be contributed by a small part of the MFR, where a partial eruption could occur that does not alter obviously the MFR as a whole.

\begin{figure}
\centering
\includegraphics[width=1.1\textwidth]{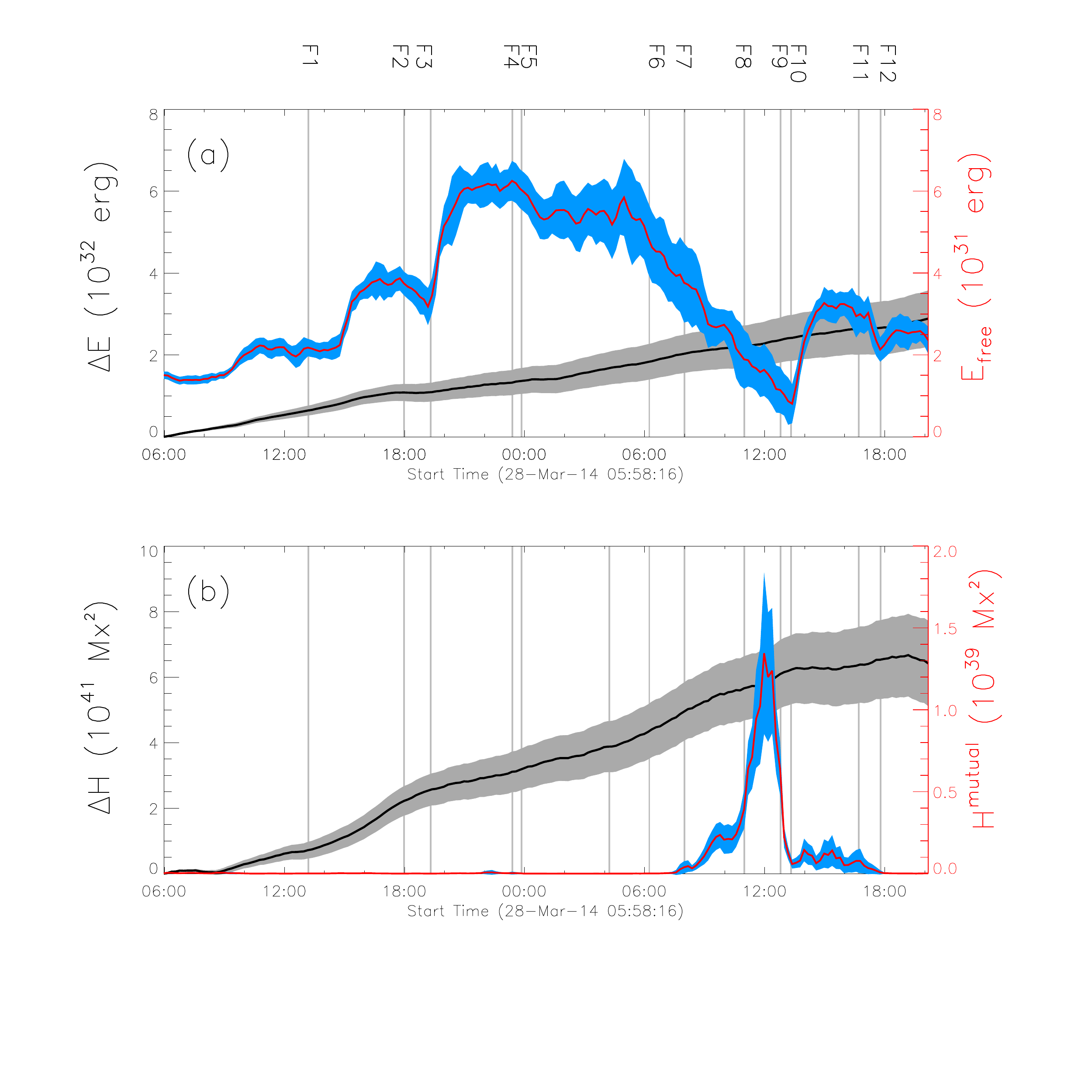} 
\caption{(a) Comparison of time integrated energy, $\Delta E$ (black line), and free energy of the MFR, $E_{free}$ (red line). (b) Comparison of time integrated helicity, $\Delta H$ (black line), and mutual helicity of the MFR, $H^{mutual}$ (red line). The vertical lines and the shaded areas have the same meaning as in Figure \ref{fig:6}.}
\label{fig:13}
\end{figure}

We also compare the magnetic energy of the MFR with the energy injected from the bottom boundary.
We find that the NLFFF energy is only less than 0.2 times the injected one and that of the potential field is even smaller.
The free energy, defined as the difference between the NLFFF energy and the potential fields energy, is only about tenth of the injected energy (Figure \ref{fig:13}(a)). 
On the other hand, unlike the total injected helicity (time integration of the helicity injection rate from the bottom boundary) that is always increasing (Figure \ref{fig:13}(b)), the magnetic helicity of the MFR does not evolve monotonically. 
Quantitatively, the latter amount to only a small fraction of the former (see also \citealt{2013Guo}). 

\subsection{Twist reduction of the MFR during confined flares}

We need to discuss more about a peculiar case, the C-class flare F9, which is associated with a large reduction of the magnetic helicity and twist of the MFR but no observed CME.
As magnetic helicity is believed to be a conserved quantity, even in resistive MHD \citep{1974Taylor}, a question thus arises where the magnetic helicity goes.
A possible explanation has been presented by \cite{2016Pinto}.
In their simulations, a confined flare is triggered by kink instability of an MFR, and magnetic reconnection occurs both within the MFR and the region between the MFR and the ambient field.
As a consequence, the twist of the MFR decreases from the initial value of $8\pi$ to the final state of $2\pi$ without a successful eruption.
It can be postulated that, in such a case, the magnetic helicity of the MFR is partly converted to that in the ambient field through magnetic reconnection.
Thus, our results for F9, with a decrease in both the twist number and magnetic helicity, and for other confined flares (F2, F4, F6, and F10), with a twist number reduction only, can be explained in terms of the above scenario.
For these events, we have confirmed that the MFR is surrounded by a QSL, which is a favourable place for magnetic reconnection \citep{1999Milano,2005Aulanier,2010WilmotSmith,2013Janvier,2009Lawrence,2012Gekelman}.
Moreover, the AIA 193 \AA \ images (Figures \ref{fig:2}) show brighting of the MFR and even above it.
In the classical flare model, post-flare loops usually appear below the MFR, as described by the CSHKP model \citep{1964Carmichael,1966Sturrock,1974Hirayama,1976Kopp}.
Thus, the AIA observations also suggest that magnetic reconnection, which is not directly related to flare loops and ribbons, occurs between the ambient field and the MFR as predicted in the simulation of \cite{2016Pinto}.

We should also mention that there is no definite decrease in the maximum twist number of the MFR during the occurrence of flares F1, F7, F8, and F11.
In some of them (e.g., F1 and F7), the maximum twist number seems to have even an increase (Table \ref{t:2}).
Possible reasons are the uncertainties in twist number calculations and the low cadence of the magnetograms.
Likewise, we also detect no significant change of the free energy during flares, except the X-class flare F12 during which the free energy decreases by about $10^{31}$ erg.
The main reason is again the errors of the free energy incurred in calculations that may reach the order of $10^{30}$ erg, a value equivalent to or larger than the total energies of some moderate and small flares.

\subsection{Summary}
\begin{figure}
\centering
\includegraphics[width=1\textwidth]{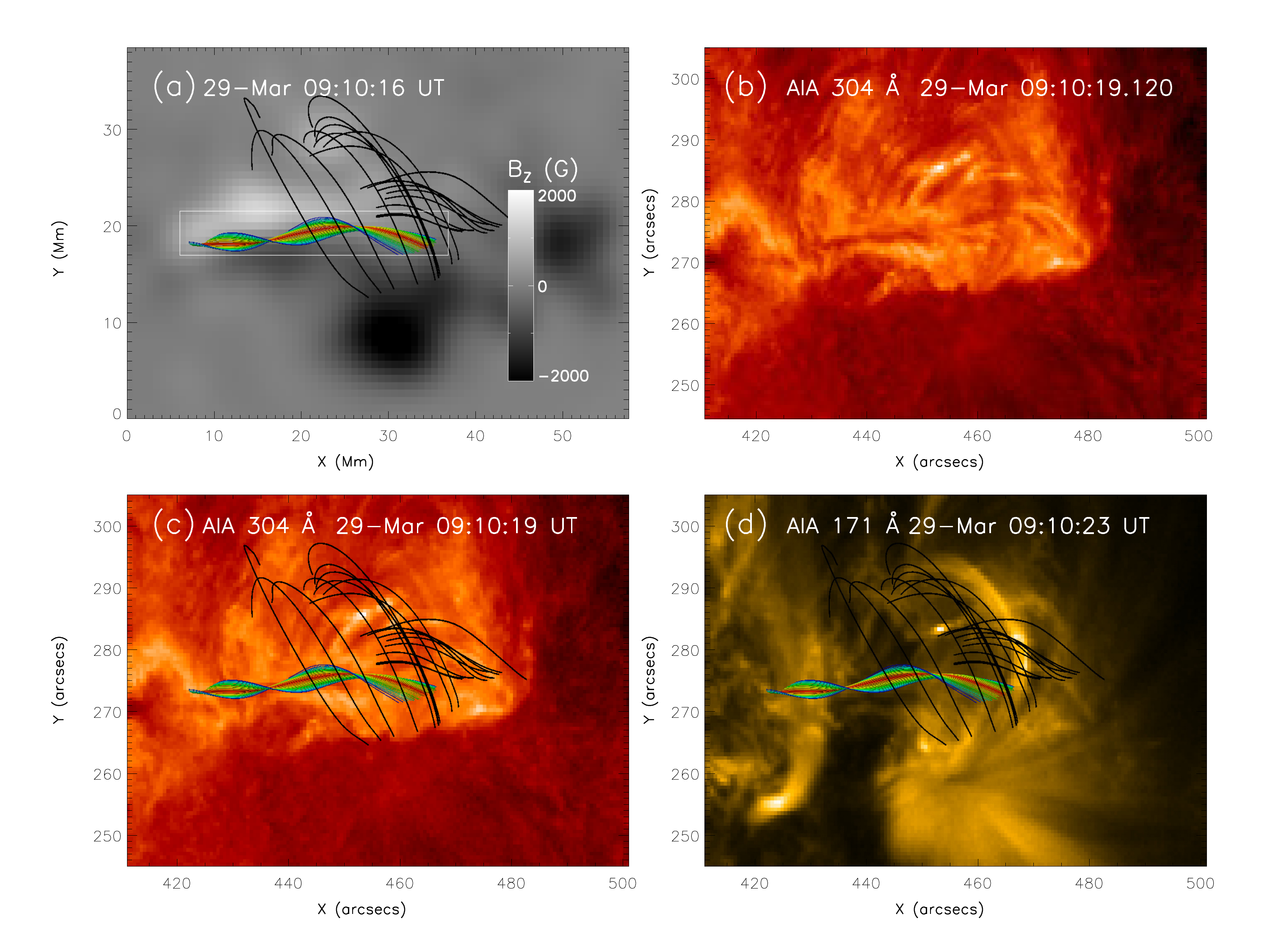} 
\caption{(a) Top view of the 3D magnetic field from the NLFFF extrapolation at 09:10:16 UT on 2014 March 29. The lines with different colors represent the MFR and the black lines represent the ambient field. The white box indicates the volume used to calculate the free magnetic energy of the MFR and the force-freeness metric near the MFR. (b) AIA 304 \AA \ image at a time close to that of panel (a). (c)--(d) Magnetic field lines overplotted on the AIA 304 \AA \ and 171 \AA \ images.}
\label{fig:7}
\end{figure}

\begin{figure}
\centering
\includegraphics[width=0.84\textwidth]{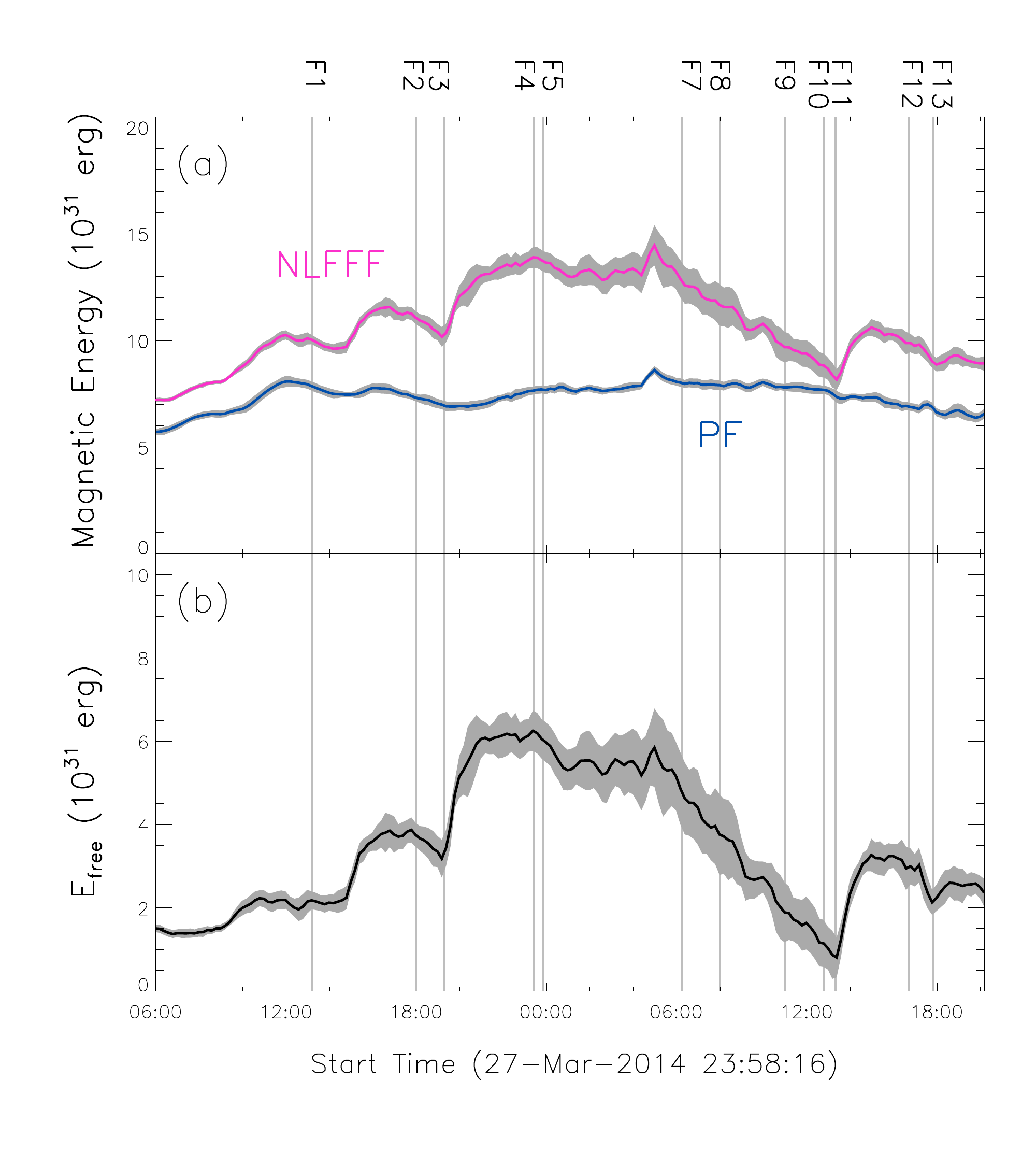} 
\caption{(a) Magnetic energy in the white box (shown in Figure \ref{fig:7}) calculated from the NLFFF extrapolation (pink) and that from the potential field (blue). (b) The free magnetic energy ($E_{free}$) contained in the white box. The vertical lines and the shaded areas have the same meaning as in Figure \ref{fig:6}.}
\label{fig:8}
\end{figure}

In this paper, we study the evolution of the MFR in a flare-productive active region 12017.
Firstly, we compute the magnetic energy and helicity injection rate corresponding to the emerging flux with the vector magnetic field coming from the data of SHARPs.
Then, we use the vector magnetic field as the boundary condition to extrapolate the 3D coronal magnetic field.
Based on the extrapolation results, an MFR is identified that is cospatial with the observed filament (Figure \ref{fig:7}).
We further compute the squashing factor around the MFR and determine the outer surface of the MFR.
The field lines threaded within the flux rope can then be obtained from the 3D magnetic field.
Finally, we compute the decay index and three quantities that are associated with the MFR: the flux-weighted average twist numbers, their associated magnetic helicities, and the mutual magnetic helicity by the method of \cite{a2006Demoulin}.
We reach the following results:
\begin{enumerate}
\item The closed QSL, enveloping the MFR, varies slowly before the flare but decreases in size sharply after the flare.
\item The flares are mainly triggered by the kink instability of the MFR that undergoes a sequence of partial eruptions.
\item The maximum twist number varies more sensitively to the flare occurrence than other parameters do.
\end{enumerate}

We should point out that the extrapolation technique assumes a force-free state of the magetic field, which may be broken when the eruptive events occur.
There are also errors from the boundary condition that include the measurement errors of the photospheric magnetic field, 180\arcdeg \ ambiguity in the horizontal field, and the projection effect.
The extrapolation results are also more or less model-dependent \citep{2012Wiegelmann,2013Regnier}.
Nevertheless, the MFR shown in the extrapolations is cospatial with the filament in AIA 304 \AA \ images, suggesting that the results are reasonable (Figure \ref{fig:7}).
In the future, we expect magnetic field measurements with higher precision and cadence that can help clarify the unclear points.

\acknowledgments
The authors thank the referee for many constructive comments that led to a significant improvement of the paper.
This work was supported by NSFC under grants 11373023 and 11203014, and NKBRSF under grant 2014CB744203.

\newpage
\bibliographystyle{apj}

\begin{thebibliography}{96}
\expandafter\ifx\csname natexlab\endcsname\relax\def\natexlab#1{#1}\fi

\bibitem[{{Amari} {et~al.}(2014){Amari}, {Canou}, \& {Aly}}]{2014Amari}
{Amari}, T., {Canou}, A., \& {Aly}, J.-J. 2014, \nat, 514, 465

\bibitem[{{Amari} {et~al.}(2000){Amari}, {Luciani}, {Mikic}, \&
  {Linker}}]{2000Amari}
{Amari}, T., {Luciani}, J.~F., {Mikic}, Z., \& {Linker}, J. 2000, \apjl, 529,
  L49

\bibitem[{{Aschwanden}(2015)}]{2015Aschwanden}
{Aschwanden}, M.~J. 2015, \apjl, 804, L20

\bibitem[{{Aulanier} {et~al.}(1999){Aulanier}, {D{\'e}moulin}, {Mein}, {van
  Driel-Gesztelyi}, {Mein}, \& {Schmieder}}]{1999Aulanier}
{Aulanier}, G., {D{\'e}moulin}, P., {Mein}, N., {van Driel-Gesztelyi}, L.,
  {Mein}, P., \& {Schmieder}, B. 1999, \aap, 342, 867

\bibitem[{{Aulanier} {et~al.}(2005){Aulanier}, {Pariat}, \&
  {D{\'e}moulin}}]{2005Aulanier}
{Aulanier}, G., {Pariat}, E., \& {D{\'e}moulin}, P. 2005, \aap, 444, 961

\bibitem[{{Aulanier} {et~al.}(2010){Aulanier}, {T{\"o}r{\"o}k}, {D{\'e}moulin},
  \& {DeLuca}}]{2010Aulanier}
{Aulanier}, G., {T{\"o}r{\"o}k}, T., {D{\'e}moulin}, P., \& {DeLuca}, E.~E.
  2010, \apj, 708, 314

\bibitem[{{Berger} \& {Field}(1984)}]{1984Berger}
{Berger}, M.~A. \& {Field}, G.~B. 1984, Journal of Fluid Mechanics, 147, 133

\bibitem[{{Berger} \& {Prior}(2006)}]{2006Berger}
{Berger}, M.~A. \& {Prior}, C. 2006, Journal of Physics A Mathematical General,
  39, 8321

\bibitem[{{Bobra} {et~al.}(2014){Bobra}, {Sun}, {Hoeksema}, {Turmon}, {Liu},
  {Hayashi}, {Barnes}, \& {Leka}}]{2014Bobra}
{Bobra}, M.~G., {Sun}, X., {Hoeksema}, J.~T., {Turmon}, M., {Liu}, Y.,
  {Hayashi}, K., {Barnes}, G., \& {Leka}, K.~D. 2014, \solphys, 289, 3549

\bibitem[{{Carmichael}(1964)}]{1964Carmichael}
{Carmichael}, H. 1964, NASA Special Publication, 50, 451

\bibitem[{{Cheng} {et~al.}(2014{\natexlab{a}}){Cheng}, {Ding}, {Zhang},
  {Srivastava}, {Guo}, {Chen}, \& {Sun}}]{a2014Cheng}
{Cheng}, X., {Ding}, M.~D., {Zhang}, J., {Srivastava}, A.~K., {Guo}, Y.,
  {Chen}, P.~F., \& {Sun}, J.~Q. 2014{\natexlab{a}}, \apjl, 789, L35

\bibitem[{{Cheng} {et~al.}(2014{\natexlab{b}}){Cheng}, {Ding}, {Zhang}, {Sun},
  {Guo}, {Wang}, {Kliem}, \& {Deng}}]{b2014Cheng}
{Cheng}, X., {Ding}, M.~D., {Zhang}, J., {Sun}, X.~D., {Guo}, Y., {Wang},
  Y.~M., {Kliem}, B., \& {Deng}, Y.~Y. 2014{\natexlab{b}}, \apj, 789, 93

\bibitem[{{Cheng} {et~al.}(2013){Cheng}, {Zhang}, {Ding}, {Liu}, \&
  {Poomvises}}]{2013Cheng}
{Cheng}, X., {Zhang}, J., {Ding}, M.~D., {Liu}, Y., \& {Poomvises}, W. 2013,
  \apj, 763, 43

\bibitem[{{Cheng} {et~al.}(2012){Cheng}, {Zhang}, {Saar}, \&
  {Ding}}]{2012Cheng}
{Cheng}, X., {Zhang}, J., {Saar}, S.~H., \& {Ding}, M.~D. 2012, \apj, 761, 62

\bibitem[{{D{\'e}moulin}(2006)}]{b2006Demoulin}
{D{\'e}moulin}, P. 2006, Advances in Space Research, 37, 1269

\bibitem[{{D{\'e}moulin} \& {Aulanier}(2010)}]{2010Demoulin}
{D{\'e}moulin}, P. \& {Aulanier}, G. 2010, \apj, 718, 1388

\bibitem[{{D{\'e}moulin} {et~al.}(1997){D{\'e}moulin}, {Bagala}, {Mandrini},
  {Henoux}, \& {Rovira}}]{1997Demoulin}
{D{\'e}moulin}, P., {Bagala}, L.~G., {Mandrini}, C.~H., {Henoux}, J.~C., \&
  {Rovira}, M.~G. 1997, \aap, 325, 305

\bibitem[{{D{\'e}moulin} \& {Berger}(2003)}]{2003Demoulin}
{D{\'e}moulin}, P. \& {Berger}, M.~A. 2003, \solphys, 215, 203

\bibitem[{{D{\'e}moulin} {et~al.}(1996{\natexlab{a}}){D{\'e}moulin}, {Henoux},
  {Priest}, \& {Mandrini}}]{a1996Demoulin}
{D{\'e}moulin}, P., {Henoux}, J.~C., {Priest}, E.~R., \& {Mandrini}, C.~H.
  1996{\natexlab{a}}, \aap, 308, 643

\bibitem[{{D{\'e}moulin} {et~al.}(2006){D{\'e}moulin}, {Pariat}, \&
  {Berger}}]{a2006Demoulin}
{D{\'e}moulin}, P., {Pariat}, E., \& {Berger}, M.~A. 2006, \solphys, 233, 3

\bibitem[{{D{\'e}moulin} {et~al.}(1996{\natexlab{b}}){D{\'e}moulin}, {Priest},
  \& {Lonie}}]{b1996Demoulin}
{D{\'e}moulin}, P., {Priest}, E.~R., \& {Lonie}, D.~P. 1996{\natexlab{b}},
  \jgr, 101, 7631

\bibitem[{{Fan}(2005)}]{2005Fan}
{Fan}, Y. 2005, \apj, 630, 543

\bibitem[{{Fan} \& {Gibson}(2003)}]{2003Fan}
{Fan}, Y. \& {Gibson}, S.~E. 2003, \apjl, 589, L105

\bibitem[{{Frisch} {et~al.}(1975){Frisch}, {Pouquet}, {Leorat}, \&
  {Mazure}}]{1975Frisch}
{Frisch}, U., {Pouquet}, A., {Leorat}, J., \& {Mazure}, A. 1975, Journal of
  Fluid Mechanics, 68, 769

\bibitem[{{Gary} \& {Hagyard}(1990)}]{1990Gary}
{Gary}, G.~A. \& {Hagyard}, M.~J. 1990, \solphys, 126, 21

\bibitem[{{Gekelman} {et~al.}(2012){Gekelman}, {Lawrence}, \& {Van
  Compernolle}}]{2012Gekelman}
{Gekelman}, W., {Lawrence}, E., \& {Van Compernolle}, B. 2012, \apj, 753, 131

\bibitem[{{Georgoulis} {et~al.}(2012){Georgoulis}, {Tziotziou}, \&
  {Raouafi}}]{2012Georgoulis}
{Georgoulis}, M.~K., {Tziotziou}, K., \& {Raouafi}, N.-E. 2012, \apj, 759, 1

\bibitem[{{Gibson} \& {Fan}(2006)}]{2006Gibson}
{Gibson}, S.~E. \& {Fan}, Y. 2006, \apjl, 637, L65

\bibitem[{{Guo} {et~al.}(2013){Guo}, {Ding}, {Cheng}, {Zhao}, \&
  {Pariat}}]{2013Guo}
{Guo}, Y., {Ding}, M.~D., {Cheng}, X., {Zhao}, J., \& {Pariat}, E. 2013, \apj,
  779, 157

\bibitem[{{Guo} {et~al.}(2012){Guo}, {Ding}, {Schmieder}, {D{\'e}moulin}, \&
  {Li}}]{2012Guo}
{Guo}, Y., {Ding}, M.~D., {Schmieder}, B., {D{\'e}moulin}, P., \& {Li}, H.
  2012, \apj, 746, 17

\bibitem[{{Guo} {et~al.}(2010){Guo}, {Ding}, {Schmieder}, {Li},
  {T{\"o}r{\"o}k}, \& {Wiegelmann}}]{2010Guo}
{Guo}, Y., {Ding}, M.~D., {Schmieder}, B., {Li}, H., {T{\"o}r{\"o}k}, T., \&
  {Wiegelmann}, T. 2010, \apjl, 725, L38

\bibitem[{{Hirayama}(1974)}]{1974Hirayama}
{Hirayama}, T. 1974, \solphys, 34, 323

\bibitem[{{Hood} \& {Priest}(1979)}]{1979Hood}
{Hood}, A.~W. \& {Priest}, E.~R. 1979, \solphys, 64, 303

\bibitem[{{Hood} \& {Priest}(1981)}]{1981Hood}
---. 1981, Geophysical and Astrophysical Fluid Dynamics, 17, 297

\bibitem[{{Janvier} {et~al.}(2013){Janvier}, {Aulanier}, {Pariat}, \&
  {D{\'e}moulin}}]{2013Janvier}
{Janvier}, M., {Aulanier}, G., {Pariat}, E., \& {D{\'e}moulin}, P. 2013, \aap,
  555, A77

\bibitem[{{Jing} {et~al.}(2012){Jing}, {Park}, {Liu}, {Lee}, {Wiegelmann},
  {Xu}, {Deng}, \& {Wang}}]{2012Jing}
{Jing}, J., {Park}, S.-H., {Liu}, C., {Lee}, J., {Wiegelmann}, T., {Xu}, Y.,
  {Deng}, N., \& {Wang}, H. 2012, \apjl, 752, L9

\bibitem[{{Judge} {et~al.}(2014){Judge}, {Kleint}, {Donea}, {Sainz Dalda}, \&
  {Fletcher}}]{2014Judge}
{Judge}, P.~G., {Kleint}, L., {Donea}, A., {Sainz Dalda}, A., \& {Fletcher}, L.
  2014, \apj, 796, 85

\bibitem[{{Kleint} {et~al.}(2015){Kleint}, {Battaglia}, {Reardon}, {Sainz
  Dalda}, {Young}, \& {Krucker}}]{2015Kleint}
{Kleint}, L., {Battaglia}, M., {Reardon}, K., {Sainz Dalda}, A., {Young},
  P.~R., \& {Krucker}, S. 2015, \apj, 806, 9

\bibitem[{{Kliem} \& {T{\"o}r{\"o}k}(2006)}]{2006Kliem}
{Kliem}, B. \& {T{\"o}r{\"o}k}, T. 2006, Physical Review Letters, 96, 255002

\bibitem[{{Kopp} \& {Pneuman}(1976)}]{1976Kopp}
{Kopp}, R.~A. \& {Pneuman}, G.~W. 1976, \solphys, 50, 85

\bibitem[{{Kusano} {et~al.}(1995){Kusano}, {Suzuki}, \&
  {Nishikawa}}]{1995Kusano}
{Kusano}, K., {Suzuki}, Y., \& {Nishikawa}, K. 1995, \apj, 441, 942

\bibitem[{{Lawrence} \& {Gekelman}(2009)}]{2009Lawrence}
{Lawrence}, E.~E. \& {Gekelman}, W. 2009, Physical Review Letters, 103, 105002

\bibitem[{{Leka} {et~al.}(2009){Leka}, {Barnes}, {Crouch}, {Metcalf}, {Gary},
  {Jing}, \& {Liu}}]{2009Leka}
{Leka}, K.~D., {Barnes}, G., {Crouch}, A.~D., {Metcalf}, T.~R., {Gary}, G.~A.,
  {Jing}, J., \& {Liu}, Y. 2009, \solphys, 260, 83

\bibitem[{{Lemen} {et~al.}(2012){Lemen}, {Title}, {Akin}, {Boerner}, {Chou},
  {Drake}, {Duncan}, {Edwards}, {Friedlaender}, {Heyman}, {Hurlburt}, {Katz},
  {Kushner}, {Levay}, {Lindgren}, {Mathur}, {McFeaters}, {Mitchell}, {Rehse},
  {Schrijver}, {Springer}, {Stern}, {Tarbell}, {Wuelser}, {Wolfson}, {Yanari},
  {Bookbinder}, {Cheimets}, {Caldwell}, {Deluca}, {Gates}, {Golub}, {Park},
  {Podgorski}, {Bush}, {Scherrer}, {Gummin}, {Smith}, {Auker}, {Jerram},
  {Pool}, {Soufli}, {Windt}, {Beardsley}, {Clapp}, {Lang}, \&
  {Waltham}}]{2012Lemen}
{Lemen}, J.~R., {Title}, A.~M., {Akin}, D.~J., {Boerner}, P.~F., {Chou}, C.,
  {Drake}, J.~F., {Duncan}, D.~W., {Edwards}, C.~G., {Friedlaender}, F.~M.,
  {Heyman}, G.~F., {Hurlburt}, N.~E., {Katz}, N.~L., {Kushner}, G.~D., {Levay},
  M., {Lindgren}, R.~W., {Mathur}, D.~P., {McFeaters}, E.~L., {Mitchell}, S.,
  {Rehse}, R.~A., {Schrijver}, C.~J., {Springer}, L.~A., {Stern}, R.~A.,
  {Tarbell}, T.~D., {Wuelser}, J.-P., {Wolfson}, C.~J., {Yanari}, C.,
  {Bookbinder}, J.~A., {Cheimets}, P.~N., {Caldwell}, D., {Deluca}, E.~E.,
  {Gates}, R., {Golub}, L., {Park}, S., {Podgorski}, W.~A., {Bush}, R.~I.,
  {Scherrer}, P.~H., {Gummin}, M.~A., {Smith}, P., {Auker}, G., {Jerram}, P.,
  {Pool}, P., {Soufli}, R., {Windt}, D.~L., {Beardsley}, S., {Clapp}, M.,
  {Lang}, J., \& {Waltham}, N. 2012, \solphys, 275, 17

\bibitem[{{Li} {et~al.}(2015){Li}, {Ding}, {Qiu}, \& {Cheng}}]{2015Li}
{Li}, Y., {Ding}, M.~D., {Qiu}, J., \& {Cheng}, J.~X. 2015, \apj, 811, 7

\bibitem[{{Liu} {et~al.}(2015){Liu}, {Deng}, {Liu}, {Lee}, {Pariat},
  {Wiegelmann}, {Liu}, {Kleint}, \& {Wang}}]{2015LiuC}
{Liu}, C., {Deng}, N., {Liu}, R., {Lee}, J., {Pariat}, {\'E}., {Wiegelmann},
  T., {Liu}, Y., {Kleint}, L., \& {Wang}, H. 2015, \apjl, 812, L19

\bibitem[{{Liu} {et~al.}(2016){Liu}, {Kliem}, {Titov}, {Chen}, {Wang}, {Wang},
  {Liu}, {Xu}, \& {Wiegelmann}}]{2016LiuR}
{Liu}, R., {Kliem}, B., {Titov}, V.~S., {Chen}, J., {Wang}, Y., {Wang}, H.,
  {Liu}, C., {Xu}, Y., \& {Wiegelmann}, T. 2016, \apj, 818, 148

\bibitem[{{Liu} {et~al.}(2012){Liu}, {Kliem}, {T{\"o}r{\"o}k}, {Liu}, {Titov},
  {Lionello}, {Linker}, \& {Wang}}]{2012Liu}
{Liu}, R., {Kliem}, B., {T{\"o}r{\"o}k}, T., {Liu}, C., {Titov}, V.~S.,
  {Lionello}, R., {Linker}, J.~A., \& {Wang}, H. 2012, \apj, 756, 59

\bibitem[{{Liu} {et~al.}(2014){Liu}, {Hoeksema}, {Bobra}, {Hayashi}, {Schuck},
  \& {Sun}}]{2014Liu}
{Liu}, Y., {Hoeksema}, J.~T., {Bobra}, M., {Hayashi}, K., {Schuck}, P.~W., \&
  {Sun}, X. 2014, \apj, 785, 13

\bibitem[{{Low}(1994)}]{1994LowBC}
{Low}, B.~C. 1994, Physics of Plasmas, 1, 1684

\bibitem[{{Mandrini} {et~al.}(1997){Mandrini}, {D{\'e}moulin}, {Bagala}, {van
  Driel-Gesztelyi}, {Henoux}, {Schmieder}, \& {Rovira}}]{1997Mandrini}
{Mandrini}, C.~H., {D{\'e}moulin}, P., {Bagala}, L.~G., {van Driel-Gesztelyi},
  L., {Henoux}, J.~C., {Schmieder}, B., \& {Rovira}, M.~G. 1997, \solphys, 174,
  229

\bibitem[{{Metcalf}(1994)}]{1994Metcalf}
{Metcalf}, T.~R. 1994, \solphys, 155, 235

\bibitem[{{Metcalf} {et~al.}(2006){Metcalf}, {Leka}, {Barnes}, {Lites},
  {Georgoulis}, {Pevtsov}, {Balasubramaniam}, {Gary}, {Jing}, {Li}, {Liu},
  {Wang}, {Abramenko}, {Yurchyshyn}, \& {Moon}}]{2006Metcalf}
{Metcalf}, T.~R., {Leka}, K.~D., {Barnes}, G., {Lites}, B.~W., {Georgoulis},
  M.~K., {Pevtsov}, A.~A., {Balasubramaniam}, K.~S., {Gary}, G.~A., {Jing}, J.,
  {Li}, J., {Liu}, Y., {Wang}, H.~N., {Abramenko}, V., {Yurchyshyn}, V., \&
  {Moon}, Y.-J. 2006, \solphys, 237, 267

\bibitem[{{Milano} {et~al.}(1999){Milano}, {Dmitruk}, {Mandrini}, {G{\'o}mez},
  \& {D{\'e}moulin}}]{1999Milano}
{Milano}, L.~J., {Dmitruk}, P., {Mandrini}, C.~H., {G{\'o}mez}, D.~O., \&
  {D{\'e}moulin}, P. 1999, \apj, 521, 889

\bibitem[{{Myers} {et~al.}(2015){Myers}, {Yamada}, {Ji}, {Yoo}, {Fox},
  {Jara-Almonte}, {Savcheva}, \& {Deluca}}]{2015Myers}
{Myers}, C.~E., {Yamada}, M., {Ji}, H., {Yoo}, J., {Fox}, W., {Jara-Almonte},
  J., {Savcheva}, A., \& {Deluca}, E.~E. 2015, \nat, 528, 526

\bibitem[{{Olmedo} \& {Zhang}(2010)}]{2010Olmedo}
{Olmedo}, O. \& {Zhang}, J. 2010, \apj, 718, 433

\bibitem[{{Pariat} \& {D{\'e}moulin}(2012)}]{2012Pariat}
{Pariat}, E. \& {D{\'e}moulin}, P. 2012, \aap, 541, A78

\bibitem[{{Pariat} {et~al.}(2005){Pariat}, {D{\'e}moulin}, \&
  {Berger}}]{2005Pariat}
{Pariat}, E., {D{\'e}moulin}, P., \& {Berger}, M.~A. 2005, \aap, 439, 1191

\bibitem[{{Park} {et~al.}(2010{\natexlab{a}}){Park}, {Chae}, {Jing}, {Tan}, \&
  {Wang}}]{a2010Park}
{Park}, S.-H., {Chae}, J., {Jing}, J., {Tan}, C., \& {Wang}, H.
  2010{\natexlab{a}}, \apj, 720, 1102

\bibitem[{{Park} {et~al.}(2010{\natexlab{b}}){Park}, {Chae}, \&
  {Wang}}]{b2010Park}
{Park}, S.-h., {Chae}, J., \& {Wang}, H. 2010{\natexlab{b}}, \apj, 718, 43

\bibitem[{{Park} {et~al.}(2013){Park}, {Kusano}, {Cho}, {Chae}, {Bong},
  {Kumar}, {Park}, {Kim}, \& {Park}}]{2013Park}
{Park}, S.-H., {Kusano}, K., {Cho}, K.-S., {Chae}, J., {Bong}, S.-C., {Kumar},
  P., {Park}, S.-Y., {Kim}, Y.-H., \& {Park}, Y.-D. 2013, \apj, 778, 13

\bibitem[{{Park} {et~al.}(2008){Park}, {Lee}, {Choe}, {Chae}, {Jeong}, {Yang},
  {Jing}, \& {Wang}}]{2008Park}
{Park}, S.-H., {Lee}, J., {Choe}, G.~S., {Chae}, J., {Jeong}, H., {Yang}, G.,
  {Jing}, J., \& {Wang}, H. 2008, \apj, 686, 1397

\bibitem[{{Pinto} {et~al.}(2016){Pinto}, {Gordovskyy}, {Browning}, \&
  {Vilmer}}]{2016Pinto}
{Pinto}, R.~F., {Gordovskyy}, M., {Browning}, P.~K., \& {Vilmer}, N. 2016,
  \aap, 585, A159

\bibitem[{{Pouquet} {et~al.}(1976){Pouquet}, {Frisch}, \&
  {Leorat}}]{1976Pouquet}
{Pouquet}, A., {Frisch}, U., \& {Leorat}, J. 1976, Journal of Fluid Mechanics,
  77, 321

\bibitem[{{Priest} \& {D{\'e}moulin}(1995)}]{1995Priest}
{Priest}, E.~R. \& {D{\'e}moulin}, P. 1995, \jgr, 100, 23443

\bibitem[{{R{\'e}gnier}(2013)}]{2013Regnier}
{R{\'e}gnier}, S. 2013, \solphys, 288, 481

\bibitem[{{Rudenko} \& {Myshyakov}(2011)}]{2011Rudenko}
{Rudenko}, G.~V. \& {Myshyakov}, I.~I. 2011, \solphys, 270, 165

\bibitem[{{Sakurai} \& {Hagino}(2003)}]{2003Sakurai}
{Sakurai}, T. \& {Hagino}, M. 2003, Advances in Space Research, 32, 1943

\bibitem[{{Savcheva} {et~al.}(2012){Savcheva}, {Pariat}, {van Ballegooijen},
  {Aulanier}, \& {DeLuca}}]{2012Savcheva}
{Savcheva}, A., {Pariat}, E., {van Ballegooijen}, A., {Aulanier}, G., \&
  {DeLuca}, E. 2012, \apj, 750, 15

\bibitem[{{Scherrer} {et~al.}(2012){Scherrer}, {Schou}, {Bush}, {Kosovichev},
  {Bogart}, {Hoeksema}, {Liu}, {Duvall}, {Zhao}, {Title}, {Schrijver},
  {Tarbell}, \& {Tomczyk}}]{2012Scherrer}
{Scherrer}, P.~H., {Schou}, J., {Bush}, R.~I., {Kosovichev}, A.~G., {Bogart},
  R.~S., {Hoeksema}, J.~T., {Liu}, Y., {Duvall}, T.~L., {Zhao}, J., {Title},
  A.~M., {Schrijver}, C.~J., {Tarbell}, T.~D., \& {Tomczyk}, S. 2012, \solphys,
  275, 207

\bibitem[{{Schou} {et~al.}(2012){Schou}, {Scherrer}, {Bush}, {Wachter},
  {Couvidat}, {Rabello-Soares}, {Bogart}, {Hoeksema}, {Liu}, {Duvall}, {Akin},
  {Allard}, {Miles}, {Rairden}, {Shine}, {Tarbell}, {Title}, {Wolfson},
  {Elmore}, {Norton}, \& {Tomczyk}}]{2012Schou}
{Schou}, J., {Scherrer}, P.~H., {Bush}, R.~I., {Wachter}, R., {Couvidat}, S.,
  {Rabello-Soares}, M.~C., {Bogart}, R.~S., {Hoeksema}, J.~T., {Liu}, Y.,
  {Duvall}, T.~L., {Akin}, D.~J., {Allard}, B.~A., {Miles}, J.~W., {Rairden},
  R., {Shine}, R.~A., {Tarbell}, T.~D., {Title}, A.~M., {Wolfson}, C.~J.,
  {Elmore}, D.~F., {Norton}, A.~A., \& {Tomczyk}, S. 2012, \solphys, 275, 229

\bibitem[{{Schrijver} {et~al.}(2008){Schrijver}, {De Rosa}, {Metcalf},
  {Barnes}, {Lites}, {Tarbell}, {McTiernan}, {Valori}, {Wiegelmann},
  {Wheatland}, {Amari}, {Aulanier}, {D{\'e}moulin}, {Fuhrmann}, {Kusano},
  {R{\'e}gnier}, \& {Thalmann}}]{2008Schrijver}
{Schrijver}, C.~J., {De Rosa}, M.~L., {Metcalf}, T., {Barnes}, G., {Lites}, B.,
  {Tarbell}, T., {McTiernan}, J., {Valori}, G., {Wiegelmann}, T., {Wheatland},
  M.~S., {Amari}, T., {Aulanier}, G., {D{\'e}moulin}, P., {Fuhrmann}, M.,
  {Kusano}, K., {R{\'e}gnier}, S., \& {Thalmann}, J.~K. 2008, \apj, 675, 1637

\bibitem[{{Schuck}(2008)}]{2008Schuck}
{Schuck}, P.~W. 2008, \apj, 683, 1134

\bibitem[{{Shen} {et~al.}(2012){Shen}, {Liu}, \& {Su}}]{2012Shen}
{Shen}, Y., {Liu}, Y., \& {Su}, J. 2012, \apj, 750, 12

\bibitem[{{Sturrock}(1966)}]{1966Sturrock}
{Sturrock}, P.~A. 1966, \nat, 211, 695

\bibitem[{{Sun}(2013)}]{2013Sun}
{Sun}, X. 2013, ArXiv e-prints

\bibitem[{{Sun} {et~al.}(2012){Sun}, {Hoeksema}, {Liu}, {Wiegelmann},
  {Hayashi}, {Chen}, \& {Thalmann}}]{2012Sun}
{Sun}, X., {Hoeksema}, J.~T., {Liu}, Y., {Wiegelmann}, T., {Hayashi}, K.,
  {Chen}, Q., \& {Thalmann}, J. 2012, \apj, 748, 77

\bibitem[{{Taylor}(1974)}]{1974Taylor}
{Taylor}, J.~B. 1974, Physical Review Letters, 33, 1139

\bibitem[{{Thalmann} {et~al.}(2011){Thalmann}, {Inhester}, \&
  {Wiegelmann}}]{2011Thalmann}
{Thalmann}, J.~K., {Inhester}, B., \& {Wiegelmann}, T. 2011, \solphys, 272, 243

\bibitem[{{Titov} {et~al.}(2002){Titov}, {Hornig}, \&
  {D{\'e}moulin}}]{2002Titov}
{Titov}, V.~S., {Hornig}, G., \& {D{\'e}moulin}, P. 2002, Journal of
  Geophysical Research (Space Physics), 107, 1164

\bibitem[{{T{\"o}r{\"o}k} \& {Kliem}(2005)}]{2005Torok}
{T{\"o}r{\"o}k}, T. \& {Kliem}, B. 2005, \apjl, 630, L97

\bibitem[{{T{\"o}r{\"o}k} {et~al.}(2004){T{\"o}r{\"o}k}, {Kliem}, \&
  {Titov}}]{2004Torok}
{T{\"o}r{\"o}k}, T., {Kliem}, B., \& {Titov}, V.~S. 2004, \aap, 413, L27

\bibitem[{{Tziotziou} {et~al.}(2013){Tziotziou}, {Georgoulis}, \&
  {Liu}}]{2013Tziotziou}
{Tziotziou}, K., {Georgoulis}, M.~K., \& {Liu}, Y. 2013, \apj, 772, 115

\bibitem[{{Tziotziou} {et~al.}(2012){Tziotziou}, {Georgoulis}, \&
  {Raouafi}}]{2012Tziotziou}
{Tziotziou}, K., {Georgoulis}, M.~K., \& {Raouafi}, N.-E. 2012, \apjl, 759, L4

\bibitem[{{Valori} {et~al.}(2012){Valori}, {D{\'e}moulin}, \&
  {Pariat}}]{2012Valori}
{Valori}, G., {D{\'e}moulin}, P., \& {Pariat}, E. 2012, \solphys, 278, 347

\bibitem[{{Wheatland} {et~al.}(2000){Wheatland}, {Sturrock}, \&
  {Roumeliotis}}]{2000Wheatland}
{Wheatland}, M.~S., {Sturrock}, P.~A., \& {Roumeliotis}, G. 2000, \apj, 540,
  1150

\bibitem[{{Wiegelmann}(2004)}]{2004Wiegelmann}
{Wiegelmann}, T. 2004, \solphys, 219, 87

\bibitem[{{Wiegelmann} {et~al.}(2006){Wiegelmann}, {Inhester}, \&
  {Sakurai}}]{2006Wiegelmann}
{Wiegelmann}, T., {Inhester}, B., \& {Sakurai}, T. 2006, \solphys, 233, 215

\bibitem[{{Wiegelmann} \& {Sakurai}(2012)}]{2012Wiegelmann}
{Wiegelmann}, T. \& {Sakurai}, T. 2012, Living Reviews in Solar Physics, 9, 5

\bibitem[{{Wilmot-Smith} {et~al.}(2010){Wilmot-Smith}, {Pontin}, \&
  {Hornig}}]{2010WilmotSmith}
{Wilmot-Smith}, A.~L., {Pontin}, D.~I., \& {Hornig}, G. 2010, \aap, 516, A5

\bibitem[{{Yang} {et~al.}(2015){Yang}, {Guo}, \& {Ding}}]{2015Yang}
{Yang}, K., {Guo}, Y., \& {Ding}, M.~D. 2015, \apj, 806, 171

\bibitem[{{Yang} {et~al.}(2013){Yang}, {B{\"u}chner}, {Santos}, \&
  {Zhang}}]{2013Yang}
{Yang}, S., {B{\"u}chner}, J., {Santos}, J.~C., \& {Zhang}, H. 2013, \solphys,
  283, 369

\bibitem[{{Young} {et~al.}(2015){Young}, {Tian}, \& {Jaeggli}}]{2015Young}
{Young}, P.~R., {Tian}, H., \& {Jaeggli}, S. 2015, \apj, 799, 218

\bibitem[{{Zhang} {et~al.}(2012){Zhang}, {Cheng}, \& {Ding}}]{2012Zhang}
{Zhang}, J., {Cheng}, X., \& {Ding}, M.-D. 2012, Nature Communications, 3, 747

\bibitem[{{Zhang} {et~al.}(2006){Zhang}, {Flyer}, \& {Low}}]{2006Zhang}
{Zhang}, M., {Flyer}, N., \& {Low}, B.~C. 2006, \apj, 644, 575

\bibitem[{{Zhang} {et~al.}(2015){Zhang}, {Ning}, {Guo}, {Zhou}, {Cheng}, {Ji},
  {Feng}, \& {Wiegelmann}}]{2015Zhang}
{Zhang}, Q.~M., {Ning}, Z.~J., {Guo}, Y., {Zhou}, T.~H., {Cheng}, X., {Ji},
  H.~S., {Feng}, L., \& {Wiegelmann}, T. 2015, \apj, 805, 4

\end{thebibliography}


\begin{deluxetable}{lccccc}
\centering
\tablewidth{0pt} 
\tablecolumns{6} 
\tablecaption{A list of the 12 flares associated with the filament in AR 12017\tablenotemark{a}\label{t:1}}
\tablehead{
\colhead{Flare label} & \colhead{Peak Time} & \colhead{} & \colhead{Class} & \colhead{CME\tablenotemark{b}} & \colhead{Active Region}}
\startdata
F1 & 28 Mar 13:12          				&        & C1.1              & no    &  12017   \\
Fb1\tablenotemark{c} & 28 Mar 16:10      &        & C1.0              & no    &  --      \\
Fb2\tablenotemark{c} & 28 Mar 17:41      &        & C1.2              & no    & 	--	    \\
F2 & 28 Mar 17:59          				&        & C2.3              & no    & 	12017   \\
F3 & 28 Mar 19:18         			    &        & M2.0              & yes   & 	12017   \\ 
Fb3\tablenotemark{c} & 28 Mar 21:00      &        & C1.0             & no     &	--	    \\ 
F4 & 28 Mar 23:24      				    &        & C1.0             & no     &	12017   \\
F5 & 28 Mar 23:51                        &        & M2.6             & yes    & 	12017\tablenotemark{d}	    \\
Fb4\tablenotemark{c} & 29 Mar 02:11	    &  		  & C2.5				 & yes 	 &	--		\\
Fb5\tablenotemark{c} & 29 Mar 04:14	    &  		  & C1.0				 & no 	 &	--		\\
Fb6\tablenotemark{c} & 29 Mar 05:02	    &  		  & C1.0				 & no 	 &	--		\\
F6 & 29 Mar 06:14               		    &        & C1.4              & no    &  12017	\\
F7 & 29 Mar 08:00           			    &        & C2.1              & no    &  12017	\\
Fb7\tablenotemark{c} & 29 Mar 10:00      & 		  & C1.4				 & no	 &	12017	\\
F8 & 29 Mar 10:59          			    &        & C1.7              & no    & 	12017	\\
Fb8\tablenotemark{c} & 29 Mar 11:34      & 		  & C1.2				 & no	 & 	12023	\\
F9 & 29 Mar 12:48                        &        & C1.5              & no   &	12017	\\
F10 & 29 Mar 13:19             			&  		  & C1.0				 & no    &	12017	\\
Fb9\tablenotemark{c} & 29 Mar 14:32      & 		  & C3.3				 & no	&	12017 	\\
F11 & 29 Mar 16:24						&  		  & C1.1				 & no   &	12017	\\
F12 & 29 Mar 17:48				        &        & X1.0              & yes  &   12017	\\
Fb10\tablenotemark{c} & 29 Mar 19:42		&  		  & C1.1				 & no   &	--		\\
Fb11\tablenotemark{c} & 29 Mar 22:48		&  		  & C2.1				 & no   &	--		\\
\enddata
\tablenotetext{a}{Flare parameters from the \textit{GOES} soft X-ray flare list.}
\tablenotetext{b}{CMEs that are checked by combining the observations of AIA and LASCO.}
\tablenotetext{c}{Flares not associated with the filament that we are interested in.}
\tablenotetext{d}{The active region hosting F5 is not marked in the \textit{GOES} website but confirmed by AIA observations.}
\end{deluxetable}


\begin{deluxetable}{lccccc}
\centering
\tablewidth{0pt} 
\tablecolumns{6} 
\tablecaption{Changes\tablenotemark{a} of the twist number during each flare.\label{t:2}}
\tablehead{
\colhead{Flare label} & \colhead{$\Delta T_{1,max}$($\%$)} & \colhead{$\Delta T_{2,max}$($\%$)} & \colhead{$\Delta\langle T_{1}\rangle$($\%$)} & \colhead{$\Delta\langle T_{2}\rangle$($\%$)}}
\startdata
F1 &   -1  $\pm$ 6   &   6   $\pm$ 5   &     5   $\pm$ 9   &    8   $\pm$ 11   \\
F2 &   -65 $\pm$ 1   &   -40 $\pm$ 3   &     -1  $\pm$ 3   &    4   $\pm$ 11   \\
F3 &   -24 $\pm$ 3   &   -22 $\pm$ 6   &     3   $\pm$ 5   &    5   $\pm$ 14   \\
F4 &   -9  $\pm$ 5   &   -7  $\pm$ 6   &     -19 $\pm$ 6   &    -25 $\pm$ 11   \\
F5 &   -32 $\pm$ 4   &   -23 $\pm$ 5   &     -25 $\pm$ 5   &    -33 $\pm$ 11   \\
F6 &   -9  $\pm$ 6   &   -13 $\pm$ 6   &     -8  $\pm$ 9   &    -12 $\pm$ 13   \\
F7 &   2   $\pm$ 5   &   7   $\pm$ 4   &     2   $\pm$ 7   &    14  $\pm$ 10   \\
F8 &   -4  $\pm$ 3   &   -6  $\pm$ 5   &     -2  $\pm$ 4   &    -2  $\pm$ 7   \\
F9 &   -15 $\pm$ 2   &   -14 $\pm$ 3   &     -10 $\pm$ 3   &    -11 $\pm$ 7   \\
F10 &  -18 $\pm$ 3   &   -13 $\pm$ 5   &     -13 $\pm$ 4   &    -20 $\pm$ 8   \\
F11 &  17  $\pm$ 6   &   2   $\pm$ 7   &     13  $\pm$ 8   &    12  $\pm$ 11   \\
F12 &  -51 $\pm$ 5   &   -47 $\pm$ 6   &     -33 $\pm$ 6   &    -47 $\pm$ 9   \\
\enddata
\tablenotetext{a}{Positive and negative signs represent increase and decrease, respectively.}
\end{deluxetable}

\newpage
\begin{figure}
\centering
\includegraphics[width=0.85\textwidth]{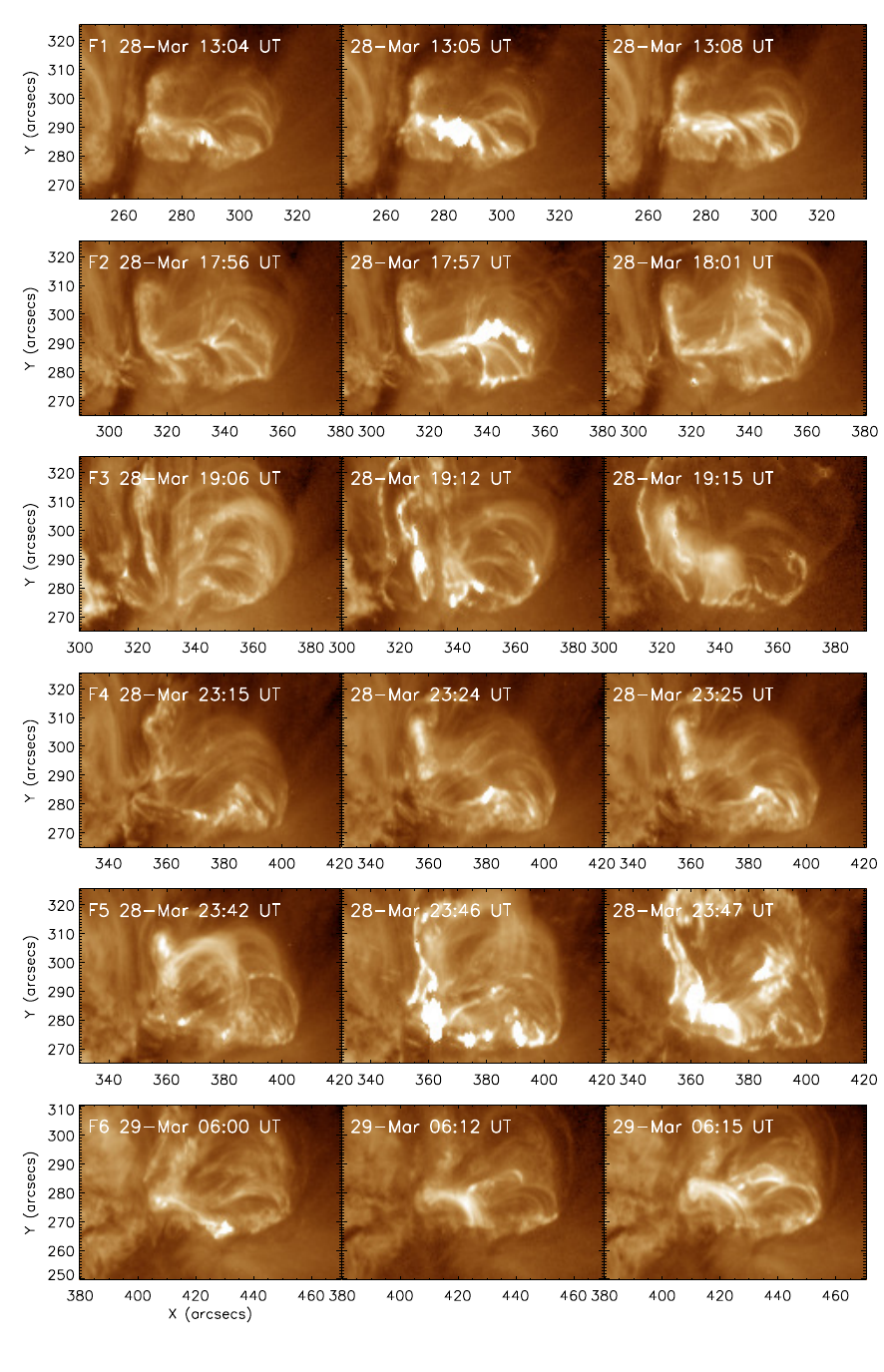} 
\caption{AIA 193 \AA \ images showing the evolution of the flares F1--F12.}
\label{fig:2}
\end{figure}
\begin{figure}
\centering
\includegraphics[width=0.85\textwidth]{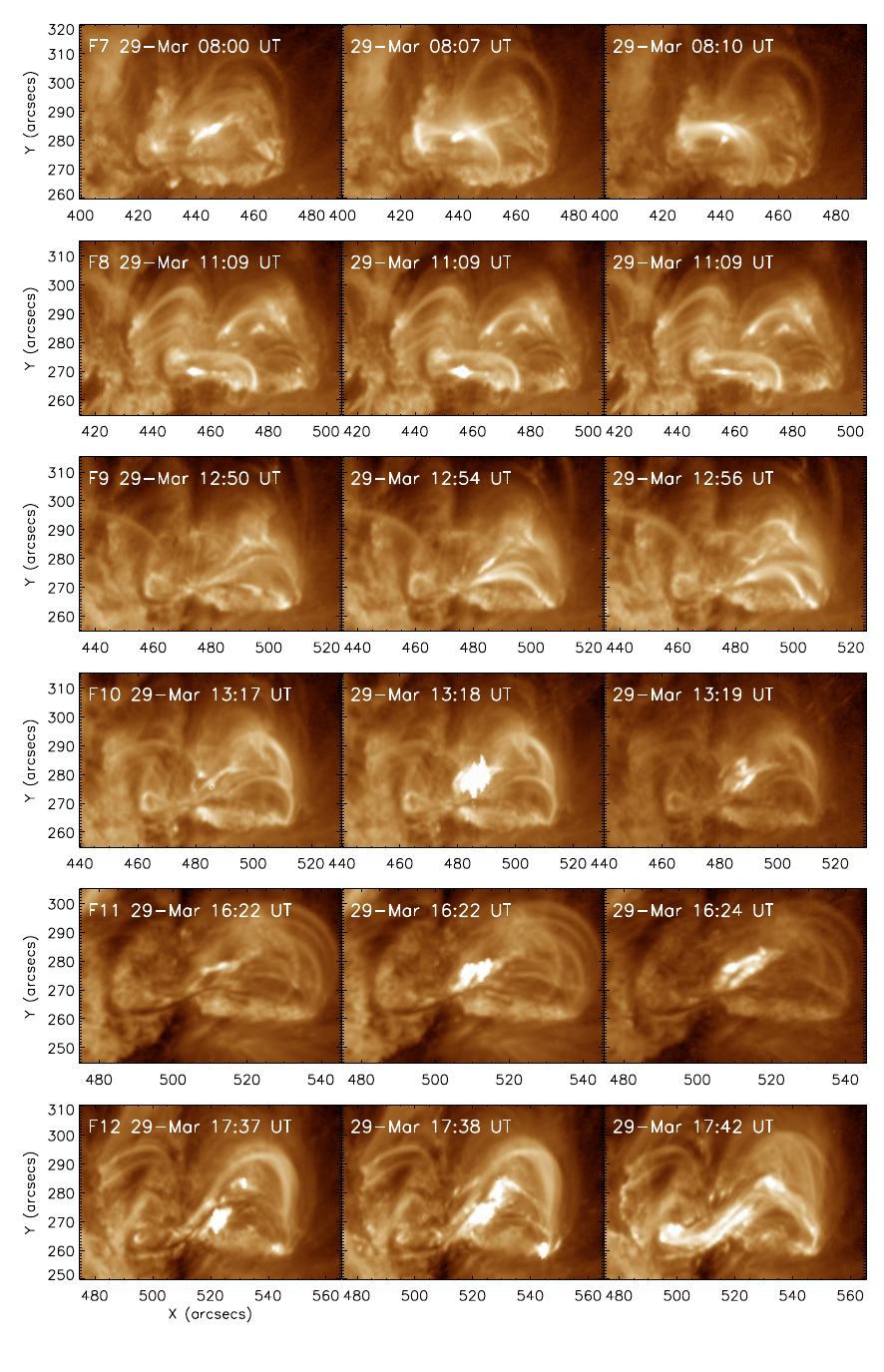} 
\figurenum{\ref{fig:2}}
\caption{Continued.}
\end{figure}
\begin{figure}
\centering
\includegraphics[width=0.7\textwidth]{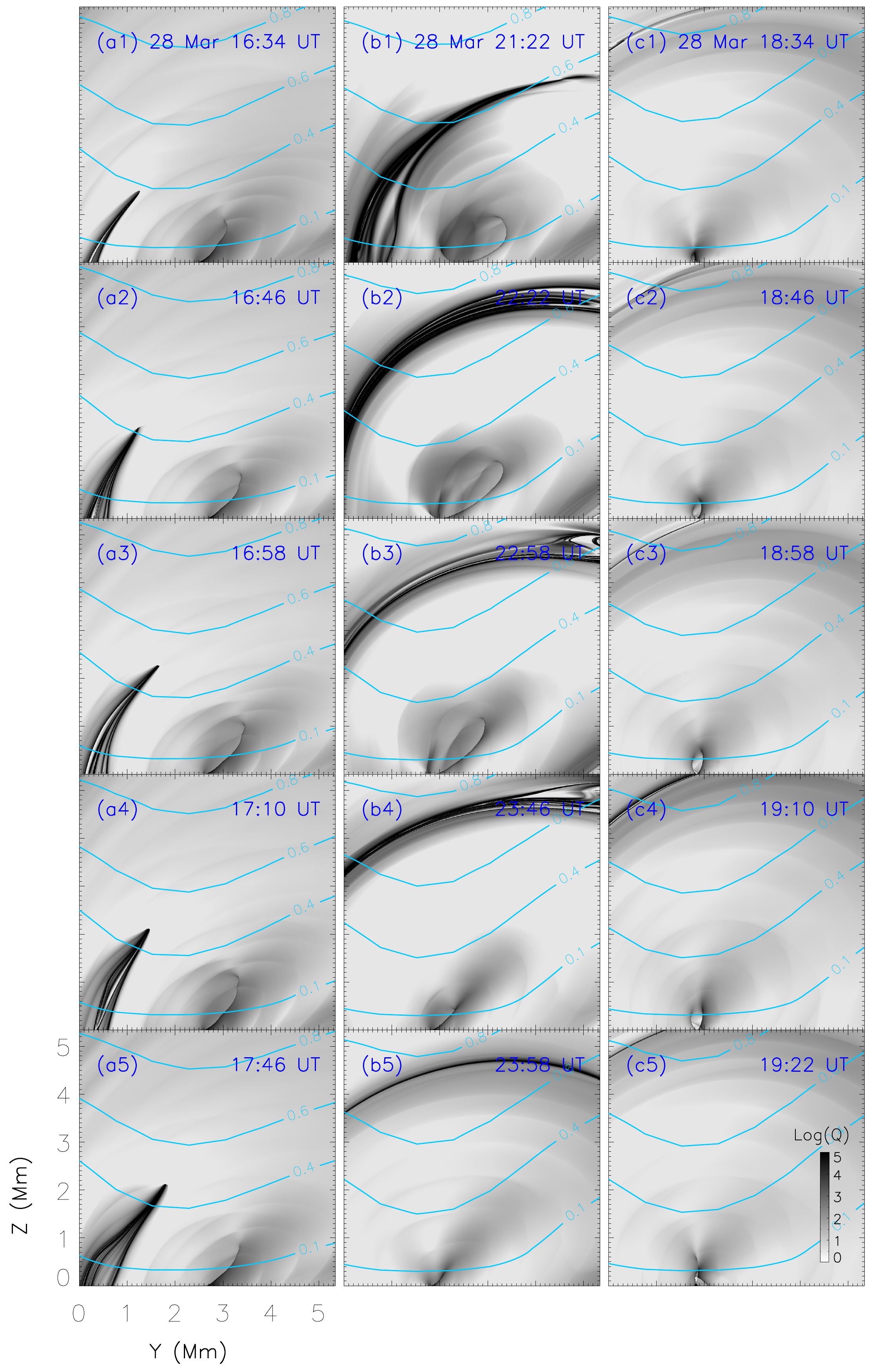} 
\caption{Distribution of the squashing factor, $Q$, on a fixed slice (shown in Figure \ref{fig:10}). Panels (a1)--(a5), (b1)--(b5), and (c1)--(c5) represent the $Q$ map evolution before (top four rows) and after (bottom row) three flares, F6, F10, and F11, respectively. Overplotted on the $Q$ map are the contours of the decay index, with the contour levels of $0.8$, $0.6$, $0.4$, and $0.1$ from top to bottom in each panel.}
\label{fig:9}
\end{figure}
\begin{figure}
\centering
\includegraphics[width=0.7\textwidth]{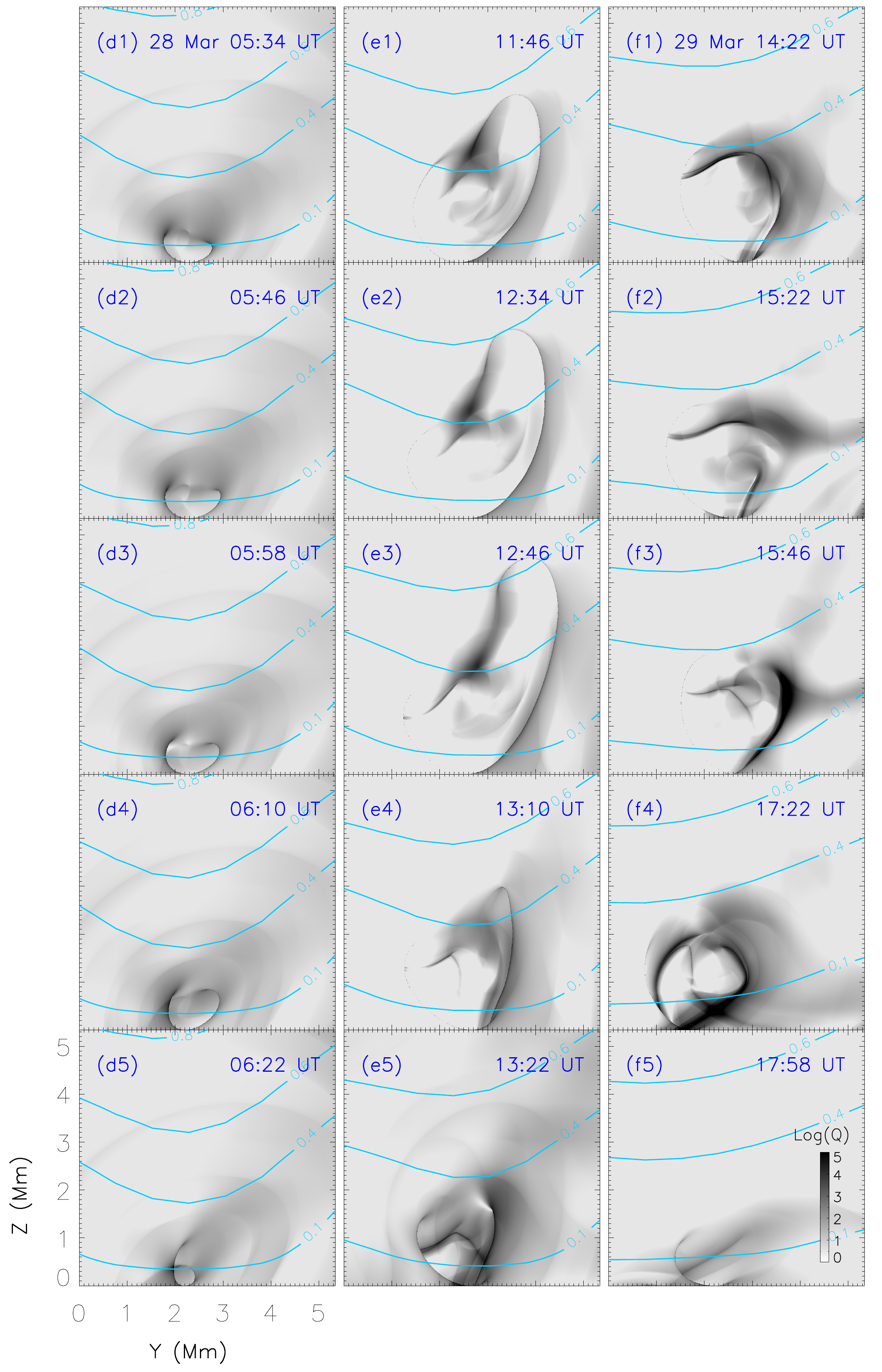} 
\figurenum{\ref{fig:9}}
\caption{Continued.}
\end{figure}
\end{document}